\documentclass[preprint,12pt]{elsarticle}

\usepackage{amssymb}
\usepackage{amsmath}

\usepackage{comment}
\usepackage{xcolor}
\usepackage[normalem]{ulem}

\newcommand{\DEL}[1]{\color{red}\sout{#1}\color{black}}

\newcommand{\mahdi}[1]{{\color{blue!50!black}#1}}
\newcommand{\saba}[1]{{\color{magenta}#1}}

\usepackage{tikz}
\usepackage{forest}
\usepackage{adjustbox}
\usepackage{array} 

\journal{In-memory Processing Architectures and Applications}

\begin{document}

\begin{frontmatter}



\title{Modeling and Simulation Frameworks for Processing-in-Memory Architectures}



\author[label1]{Mahdi Aghaei}
\author[label1]{Saba Ebrahimi}
\author[label1]{Mohammad Saleh Arafati} 
\author[label1]{Elham Cheshmikhani}
 \author[label1]{Dara Rahmati}
\affiliation[label1]{organization={Dept. of Computer Science and Engineering, Shahid Beheshti University},
            city={Tehran},
            country={Iran}}
            



 \author[label2]{Saeid Gorgin}
\author[label2]{Jungrae Kim}
\affiliation[label2]{organization={Dept. of Electrical and Computer Engineering, Sungkyunkwan University}, 
            city={Seoul},
            country={Korea}}

\begin{abstract}
Processing-in-Memory (PIM) has emerged as a promising computing paradigm to address the memory wall and the fundamental bottleneck of the von Neumann architecture by reducing costly data movement between memory and processing units. As with any engineering challenge, identifying the most effective solutions requires thorough exploration of diverse architectural proposals, device technologies, and application domains. In this context, simulation plays a critical role in enabling researchers to evaluate, compare, and refine PIM designs prior to fabrication.
Over the past decade, a variety of PIM simulators have been introduced, spanning low-level device models, architectural frameworks, and application-oriented environments. These tools differ significantly in fidelity, scalability, supported memory/compute technologies, and benchmark compatibility. Understanding these trade-offs is essential for researchers to select appropriate simulators that accurately map and validate their research efforts.

This chapter provides a comprehensive overview of PIM simulation methodologies and tools. We categorize simulators according to abstraction levels, design objectives, and evaluation metrics, highlighting representative examples. To improve accessibility, some content may appear in multiple contexts to guide readers with different backgrounds. We also survey benchmark suites commonly employed in PIM studies and discuss open challenges in simulation methodology, paving the way for more reliable, scalable, and efficient PIM modeling.

\end{abstract}

%

\end{frontmatter}




\section{Introduction}
\label{sec1}
\subsection{The Necessity and Role of Simulation}
\label{sec:necessity_simulation}
Simulation forms the methodological backbone of modern computer architecture research. The prohibitive cost, design complexity, and risks associated with fabricating experimental hardware prototypes render physical evaluation impractical for most exploratory concepts.
Simulation environments provide a controllable and reproducible framework to estimate performance, energy efficiency, and functional correctness before committing designs to silicon. Through simulation, designers can iterate rapidly on design parameters, identify performance bottlenecks, and evaluate competing architectural alternatives.

In the broader context of computing systems, reliance on simulation has grown in proportion to the complexity of hardware–software interactions. From instruction set simulators to detailed timing models, researchers employ a variety of tools to capture system behavior across different abstraction layers. These simulators help bridge the gap between high-level algorithmic modeling and low-level hardware verification enabling quantification of trade-offs among performance, area, power, and thermal metrics, well in advance of fabrication.

For emerging paradigms like Processing-in-Memory (PIM), simulation plays an even more essential role. Unlike conventional CPUs and GPUs, where architectural components are well established, PIM architectures are highly diverse and continuously evolving. Each design may integrate computation with memory arrays, logic layers adjacent to DRAM stacks, or specialized non-volatile devices. Constructing physical prototypes for every such configuration is economically unfeasible and technologically constrained. Consequently, simulation becomes the de facto tool for evaluating architectural ideas and quantifying their impact on system performance.

Moreover, simulation fosters scientific reproducibility. Published PIM architectures are often evaluated under varying workloads, device technologies, and memory hierarchies. Without standardized simulation methodologies, comparing results across studies is unreliable. Simulator frameworks provide a common foundation for validation, ensuring that future work can be consistently benchmarked prior research.

From an educational standpoint, simulation also democratizes architectural research. Graduate students and engineers can explore advanced design ideas using open-source simulators without access to costly fabrication facilities. In the industry, simulators shorten design cycles by enabling hardware teams to evaluate multiple design options in parallel. Indeed, simulation is not merely a convenience, it is a prerequisite for innovation in data-centric computing.


\subsection{Simulation in the Context of PIM}
\label{sec:intro_simulationIPM}

Processing-in-Memory (PIM) challenges the long-standing von Neumann paradigm by integrating memory and computation into a single substrate. This fusion significantly reduces data movement, the primary source of energy consumption and latency in modern workloads. However, this same integration complicates analysis. Unlike conventional CPU pipelines or standard DRAM interfaces, PIM involves simultaneous modeling of both computation and memory behaviors. Simulation therefore becomes the only practical approach for evaluating performance interactions and verifying architectural correctness.

Effective PIM simulation requires co-modeling across multiple subsystems, including the memory array, peripheral circuits, logic layer, and host interface. Each component introduces critical parameters, such as latency, energy consumption, bandwidth, and fault tolerance, that collectively shape system-level behavior. Moreover, the diversity of physical memory technologies (e.g., DRAM, SRAM, RRAM, PCM, MRAM, and hybrid 3D-stacked memories) demands flexible and extensible simulation frameworks that can adapt to diverse device characteristics.

Different abstraction levels serve different research purposes. Device-level simulators focus on electrical properties, such as resistive switching and retention. Circuit-level simulators analyze sense-amplifier timing behavior in components such as sense amplifiers, row activation, and bitline coupling. Architectural simulators estimate throughput and energy consumption across memory banks, channels, and embedded processing elements. Full-system simulators integrate CPUs, interconnects, and software stacks to provide end-to-end system insights. Together, these simulation layers enable researchers to traverse the entire design hierarchy, from device physics to system deployment, without requiring hardware.

Furthermore, PIM simulation enables algorithm–architecture co-design. Workloads such as machine-learning kernels, graph analytics, and database operations can be mapped to different PIM substrates to evaluate efficiency gains. Feedback from simulation informs hardware designers about bottlenecks such as  bandwidth limitations or thermal hotspots, prompting interactive architectural refinement. In this way, simulation functions as both a predictive and prescriptive instrument, guiding the evolution of PIM from concept to implementation.

\subsection{A Timeline of PIM Simulators}
\label{sec:timeline}

The historical progression of PIM simulators mirrors the technological evolution of memory-centric computing.  
Early frameworks were largely DRAM-centric, extending memory simulators such as \texttt{NVMain}~\cite{Poremba2012NVMain} with simplified compute modules. These tools captured bandwidth and latency but ignored computation–memory coupling.  

Between 2016 and 2020, specialized simulators like \texttt{PIMSim}~\cite{xu2018pimsim} and \linebreak \texttt{Ramulator-PIM}~\cite{git2025ramulatorPIM} introduced programmable PIM kernels and near-memory logic models. 
From 2020 onward, research shifted toward architectural and full-system~co-simulation, exemplified by \texttt{PIMSimulator}~\cite{git2025PIMSimulator} and \texttt{PiMulator}~\cite{Mosanu2022PiMulator}.

\begin{figure}[h]
\centering
\includegraphics[scale=0.35]{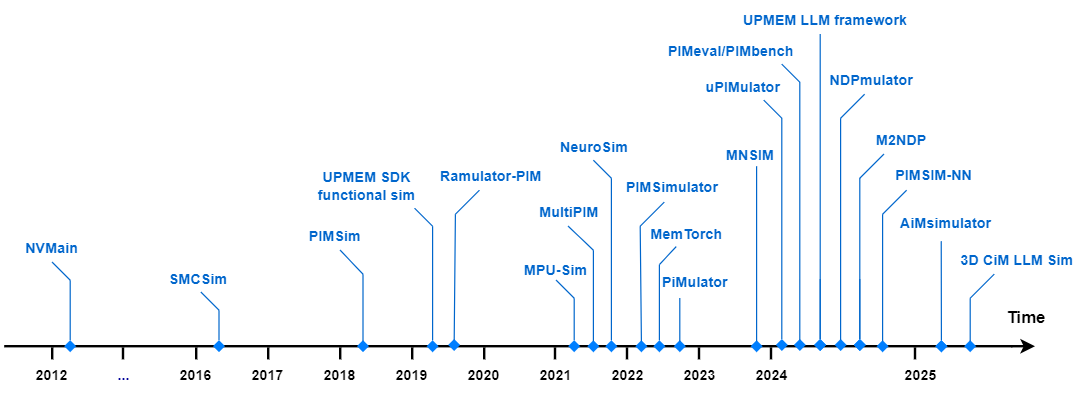}
\caption{Timeline of PIM simulators}
\label{fig:timeline}
\end{figure}

Recent frameworks, including \texttt{PIMeval}~\cite{10763591} and \texttt{MultiPIM}~\cite{Yu2021MultiPIM}, integrate benchmark suites, performance counters, and support for AI/ML workloads.  
This progression highlights a steady movement from feasibility analysis to end-to-end performance evaluation and application-driven design. Figure~\ref{fig:timeline} can illustrate this progression, showing the shift from memory-centric models to holistic, application-aware simulation frameworks.


\subsection{Practical Guide for Readers}
\label{sec:readers_guide}
\begin{figure}[t]
\centering
\includegraphics[scale=0.3]{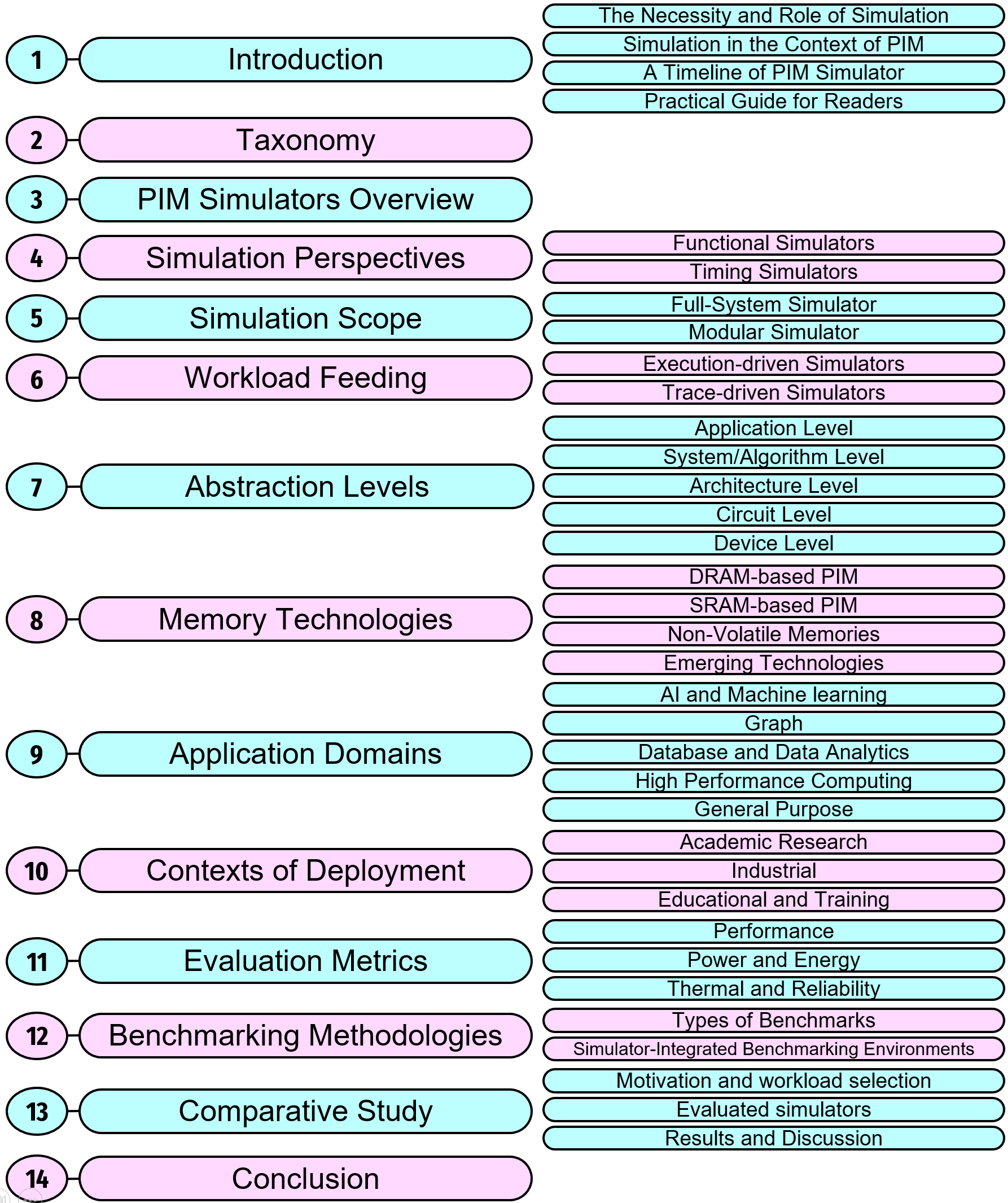}
\caption{Overall structure of this chapter}
\label{fig:chapter_structure}
\end{figure}

Given the diversity of available tools, selecting the most suitable simulator depends on the reader’s specific objective and expertise. This section provides a practical guide to help readers navigate the extensive and diverse landscape of PIM simulators. The organization of this chapter is intentionally hierarchical: early sections establish the conceptual foundation, while later sections delve into specialized simulation methodologies and case studies. The overview of the chapter’s structure is depicted in Figure~\ref{fig:chapter_structure}.
Readers are encouraged to start with the initial taxonomy and overview sections before exploring more technical ones. 

\begin{itemize}
  \item Section~\ref{sec:taxo} – Taxonomy of PIM Simulators: 
  introduces a comprehensive classification framework used throughout the book. It defines key dimensions including accuracy, scope, abstraction, technology, and evaluation metrics, that underpin the organization of subsequent sections.
  \item Section~\ref{sec:overview}  – Overview of Existing PIM Simulators:
  provides a concise survey of representative open-source simulators. It summarizes their design objectives, availability, and supported memory technologies, serving as a quick reference before readers engage with detailed methodological discussions.
  \item Sections~\ref{sec:sim_perspectives} and \ref{sec:sim_scope} – Simulation Perspectives and Scope: 
  together, these sections examine the methodological dimensions of PIM simulation. Section~\ref{sec:sim_perspectives} discusses timing fidelity and functional accuracy, distinguishing between functional, event-driven, and cycle-level methodologies. Section~\ref{sec:sim_scope} expands this discussion to system scope, comparing full-system versus modular simulation approaches and analyzing trade-offs between coverage, scalability, and integration complexity. 
  \item Section~\ref{sec:sim_workload} – Workload Feeding Mechanisms:
  details execution-driven and trace-driven methodologies used to provide stimuli to simulators.
  \item Section~\ref{sec:sim_abstractions} and \ref{sec:mem_tech} – Abstraction Levels and Memory Technologies:
  discuss the range of simulation abstraction from device physics to system architecture, and examine how various memory technologies influence simulation design.
  \item Section~\ref{sec:sim_domains} and \ref{sec:sim_deployment} – Application Domains and Deployment Context:
  present how simulators target distinct workloads (AI, graph analytic, database processing, HPC) and cater to different user communities (academic, industrial, educational).
  \item Section~\ref{sec:sim_metrics} – Evaluation Metrics for PIM Simulation: 
  synthesizes methodologies for assessing performance, energy, and reliability, and introduces benchmark suites such as PIMBench.
  \item Section~\ref{sec:sim_benchmarks} – Benchmarking Methodologies for PIM Architectures: introduces benchmark suites and standardized workloads designed specifically for PIM evaluation. It examines representative frameworks such as \texttt{PIMBench}, and their domain-tailored workloads for AI inference, graph traversal, and database acceleration.
  \item Section~\ref{sec:comparative_study} – Comparative Study of PIM Simulators: presents a quantitative evaluation of two representative open-source PIM simulation frameworks, \texttt{PIMeval} and \texttt{PIMSimulator}. It compares their modeling depth, configuration flexibility, and performance predictions across GEMM and GEMV workloads, highlighting trade-offs between functional abstraction and timing fidelity in PIM system evaluation.
  \item Section~\ref{sec:conclusion} – Conclusion: Summarizes key insights from from all previous chapters, highlighting open challenges and future research directions in PIM simulation, including fidelity scaling, integration with machine-learning-driven design-space exploration, and co-verification with hardware prototypes.
\end{itemize}

\section{Taxonomy of PIM Simulators}
\label{sec:taxo}

This section presents a compact taxonomy that organizes PIM simulators along eight orthogonal axes. Each axis captures a key design decision that researchers must consider when selecting  simulation tools, and each is expanded in detail in later sections. The goal is to provide a concise, self-contained “map” that enables readers to navigate directly to the most relevant deep-dive with a shared terminology.

\subsection{Accuracy Perspectives}
\label{subsec:accuracy}
PIM simulators differ fundamentally in how they model time. Functional simulators prioritize logical correctness and rapid iteration. They validate ISA extensions, kernel semantics, and software stacks behavior without incorporating detailed timing~\cite{Buitrago_Com_2024,8718630}. Timing simulators including models, capture dynamic behaviors such as queuing, hazards, contention, and device latencies. These are essential for quantifying execution throughput, latency, and overlap between host and PIM execution~\cite{8718630,7753351}.
In practice, researchers often follow a top–down workflow: start with functional models, for rapid prototyping and progressing to timing-accurate simulations to refine insights. Hybrid models augment functional cores with analytical delay and energy estimators, providing early visibility into performance trends while maintaining simulation speed.
Readers seeking trade-offs between speed and fidelity, and the distinctions among functional, event-driven, and cycle-accurate approaches, should see Section~\ref{sec:sim_perspectives}, where these perspectives are decomposed, exemplified, and compared with representative tools.

\subsection{Scope and Granularity}
\label{subsec:scope}
Scope describes \emph{how much of the system} a simulator models. \emph{Full-system} frameworks emulate CPUs, memory, I/O, and OS, enabling end-to-end studies of scheduling, drivers, coherence, and virtual memory interaction with PIM~\cite{10.5555/1855084}. Modular frameworks focus on specific subsystems (e.g., DRAM timing, vault controllers, or PIM kernels) for scalability and targeted microarchitectural analysis~\cite{8718630}.
Granularity also plays a role: coarse-grain models accelerate design-space sweeps by simplifying details; fine-grain ones expose bottlenecks and corner cases but are typically slower. When selecting scope and granularity, align them with your question: Prefer modular models for microarchitectural sensitivity analysis or ``what-if'' explorations of individual design components.
A structured discussion and case study appear in Section~\ref{sec:sim_scope}.

\subsection{Workload Feeding Mechanisms}
\label{subsec:tax_feeding}
How a simulator is ``fed'' determines realism, reproducibility, and cost. \emph{Execution-driven} flows run binaries inside the simulated platform, naturally exposing dynamic effects (speculation execution, runtime scheduling, contention) and eliminating massive trace files~\cite{10.5555/1855084}.
\emph{Trace-driven} simulators flow replay pre-collected instruction/memory traces, trading some dynamism for excellent repeatability and fast multi-configuration sweeps; they are ideal for broad design exploration and regression testing~\cite{8718630}. Many projects combine both approaches: generate traces with an execution-driven run, then replay across architectures or device models. For a systematic comparison, practical guidelines, and representative tools, see Section~\ref{sec:sim_workload}.

\subsection{Abstraction Levels}
\label{subsec:tax_abstraction}
PIM research spans five principal abstractions: \emph{Application} (end-to-end workload behavior and KPIs), \emph{System/Algorithm} (mapping/scheduling across host and memory-side compute), \emph{Architecture} (controllers, kernels, interconnects, coherence), \emph{RTL/Circuit} (cycle-level pipelines, per-block timing/power), and \emph{Device} (cell physics, endurance, variability)~\cite{maurer2025survey}. Choosing the right level hinges on your hypothesis and validation needs. For example, algorithm fit and speedups suggest application/system levels; ISA/coherence proposals require architectural timing; reliability/analog effects push toward circuit/device co-simulation. Cross-level coupling (e.g., architecture+device) is increasingly common for non-idealities and thermal constraints. A detailed, level-by-level treatment appears in Section~\ref{sec:sim_abstractions}.

\subsection{Memory Technologies}
\label{subsec:tax_tech}
The underlying memory technology shapes the set of feasible PIM operations, timing characteristics, and energy consumption profiles. DRAM-based PIM solutions (DDR/HBM/HMC/UPMEM) favors near-bank digital kernels and offers high internal bandwidth; SRAM-based designs trade density for ultra-low latency and tight on-chip integration; Non-volatile options (ReRAM/PCM/MRAM/FeFET) enable analog/digital in-place compute with endurance and variability trade-offs \cite{mohammadi2025reliability, cheshmikhani2024low}; emerging stacks and coherent fabrics (e.g., CXL-attached NDP) broaden integration possibilities. Selecting or abstracting the appropriate technology model is crucial to draw valid conclusions about performance, energy efficiency, and reliability. Technology-specific modeling considerations are comprehensively surveyed in Section~\ref{sec:mem_tech}.

\subsection{Application Domains of PIM Simulators}
\label{subsec:tax_domains}
PIM simulators target distinct workload classes with different access patterns and compute intensities. AI/ML emphasizes GEMM/GEMV, attention, and Key-Value (KV)-caches~\cite{10.1145/3695794.3695797}; \emph{Graph} processing workloads stresses irregular, pointer-chasing memory~\cite{ahn2015scalable};
\emph{Databases/Analytics} are dominated by scans, joins, and reductions~\cite{10.14778/3368289.3368298}; \emph{High-performance computing (HPC)} mixes dense/sparse linear algebra and stencil codes~\cite{denzler2023casper}.

Some frameworks are \emph{domain-agnostic}, offering configurable kernels and memory backends for broad cross-domain comparisons. Domain alignment is essential for credible evaluation: a simulator validated on CNN inference may not capture graph frontier behavior or OLAP joins. Representative mappings and decision criteria are provided in Section~\ref{sec:sim_domains}.

\subsection{Contexts of Deployment}
\label{subsec:tax_contexts}
The same simulator can serve different proposes across communities. \emph{Academic} settings use prioritizes extensibility, openness, and rapid prototyping for new ideas and cross-layer experiments. \emph{Industrial} contexts emphasize calibrated timing, determinism, toolchain integration, and pre-silicon verification capabilities. \emph{Educational/training} environments favor clarity, setup simplicity, and visual explainability. Recognizing these diverse deployment contexts helps set appropriate accuracy/runtime expectations and reporting standards (e.g., reproducible configs vs. silicon-correlated models). Guidance and representative tools tailored to each context appear in Section~\ref{sec:sim_deployment}.

\subsection{Evaluation Metrics for PIM Simulation}
\label{subsec:tax_metrics}
Robust quantitative assessment hinges on a balanced metric suite: \emph{performance} (latency, throughput, bandwidth efficiency), \emph{power/energy} ( TOPS/W, total energy), and \emph{thermal/reliability} (hotspots detection, fault injection, endurance/ retention)~\cite{10.5555/1855084}. Cross-simulator comparability requires transparent methodology including identical datasets, calibration points, warm-up/\linebreak measurement intervals, and statistical treatment.
A practical approach combines fast analytical models for sweeps with detailed runs for focal points, and report sensitivity to technology and mapping choices. A structured treatment, including cross-validation practices, is provided in Section ~\ref{sec:sim_metrics}.

\subsection{PIM-Specific Benchmarks}
\label{subsec:tax_bench}
Benchmarks operationalize the taxonomy by standardizing inputs, kernels, and evaluation metrics. General benchmark suites (e.g., ML/DL, graph, database, and HPC) should be complemented with \emph{PIM-specific} collections that expose near-memory kernels (reductions, scans, atomics, bitwise ops, analog MVM) and memory-stressing access patterns.

High-quality PIM benchmarks define clear input scaling rules, mapping rules, and reference baselines (e.g., CPU, GPU, or accelerator), along with harnesses capable of generating traces or executing natively on simulators. They should also document hardware and technology assumptions to ensure fair comparisons in latency and energy efficiency. A survey of available benchmark suites and their workload coverage are presented in Section~\ref{sec:sim_benchmarks}.

\section{PIM Simulators Overview}
\label{sec:overview}
This section provides a comprehensive overview of the publicly available Processing-in-Memory (PIM) simulators that have been released to the research community. The focus is deliberately restricted to frameworks whose source code is accessible, either through open-source repositories or publicly shared academic distributions, ensuring that the tools discussed here can be directly used, verified, and extended by other researchers.  
While numerous PIM-related simulators have been proposed over the past decade, many remain proprietary, inaccessible, or documented only at a high level. Such tools, though valuable for industrial or internal evaluation, cannot serve as reproducible baselines for academic investigation. Therefore, only simulators that are fully or partially open-source, accompanied by functional documentation or repositories, are included in this section.

\subsection{NVMain}
\texttt{NVMain}~\cite{Poremba2012NVMain} is a cycle-accurate, architectural-level main memory simulator developed to model both conventional DRAM and emerging non-volatile memory (NVM) technologies such as ReRAM, STT-RAM, and PCM. It was designed to address the limitations of DRAM-centric simulators (e.g., \texttt{DRAMSim}) that lack support for endurance modeling, asymmetric read/write latencies, and hybrid DRAM–NVM systems. The simulator models complete memory subsystems—including channels, ranks, banks, and subarrays—with fully customizable timing parameters (e.g., $t{RCD}$, $t{RAS}$, $t{WR}$, $t{FAW}$) and built-in verification for command legality and bus contention. Moreover, \texttt{NVMain} can import latency and energy parameters from circuit-level tools such as \texttt{NVSim} (built on \texttt{CACTI}), thereby bridging device-level modeling with architectural exploration. Its modular architecture supports experimentation with memory controllers, refresh policies, and interconnects—from traditional DDR buses to alternative optical or serial links.

In addition to timing and performance, \texttt{NVMain} incorporates detailed endurance and reliability modeling, enabling long-term studies of wear-out behavior and fault propagation in non-volatile arrays. It supports advanced techniques such as Data-Comparison Write (DCW) and Flip-N-Write to minimize redundant bit transitions, improving both write energy efficiency and device lifetime. Power estimation can rely on IDD-based DRAM power models or on technology-specific data imported from \texttt{NVSim}/\texttt{CACTI}, while endurance-aware policies and refresh strategies can be evaluated under realistic workload traces.

Due to its extensibility and cycle-level precision, \texttt{NVMain} has become a foundational component in subsequent PIM-related simulators such as \texttt{PIMSim}, \texttt{MultiPIM}, and \texttt{NDPmulator}. These frameworks reuse \texttt{NVMain}'s validated energy and timing models as their back-end memory substrate to support heterogeneous DRAM–NVM configurations and to analyze data-movement efficiency in near-memory computing systems. Overall, \texttt{NVMain} remains one of the most influential open-source simulators for bridging the gap between circuit-level device models and system-level architectural evaluation of next-generation memory technologies.

\subsection{NVSim}
\texttt{NVSim}~\cite{Dong2012NVSim} is a circuit-level modeling framework designed to estimate the area, latency, dynamic energy, and leakage power of both volatile and nonvolatile memory (NVM) technologies. Supporting a wide spectrum of emerging memories—such as Phase-Change Memory (PCM/PCRAM), Spin-Transfer Torque RAM (STT-RAM/MRAM), Resistive RAM (ReRAM), Floating-Body DRAM (FBDRAM), and NAND Flash—\texttt{NVSim} provides a unified analytical platform for device-to-architecture performance evaluation.

Developed as an extension of the modeling concepts introduced by \texttt{CACTI}, \texttt{NVSim} emphasizes configurability and extensibility, offering detailed parameterization of banks, mats, subarrays, and peripheral circuits. Its framework accommodates diverse routing topologies (e.g., H-tree, bus-based architectures) and multiple sensing schemes—current-mode, voltage-mode, and voltage-divider—allowing accurate capture of technology-specific read and write behaviors. Furthermore, it integrates process-dependent parameters, such as resistance range, write current, and retention characteristics, to support exploration of novel NVM materials and device structures.

Validation results demonstrate that \texttt{NVSim} achieves high modeling fidelity, typically within 30\% of empirical measurements from industrial prototypes. While originally developed for conventional memory design rather than Processing-in-Memory (PIM), \texttt{NVSim} has become a foundational component for higher-level PIM and Compute-in-Memory (CIM) simulators. Many later tools—such as \texttt{NVSim-PIM}, \texttt{NVMain}, and \texttt{NeuroSim}—inherit its timing and power estimation modules to enable more complex architectural and system-level modeling.

In essence, \texttt{NVSim} functions as a functional, modular, and circuit-level simulator that facilitates early-stage design-space exploration of emerging memory technologies. Although it does not provide full-system or cycle-accurate simulation, its ability to bridge device-level parameters with architectural-level insights has made it a cornerstone in the evaluation and optimization of both standalone NVM designs and memory-centric computing architectures.

\subsection {SMCSim}
\texttt{SMCSim}~\cite{Azarkhish2016SMCSim} is a full-system, gem5-based simulation framework developed to model and evaluate near-memory computation within the Smart Memory Cube (SMC)—an enhanced evolution of the Hybrid Memory Cube (HMC) that integrates programmable compute units in its logic base (LoB). Built atop gem5’s General Memory System infrastructure, \texttt{SMCSim} reuses fundamental components such as DMA engines, interconnect fabrics, and DRAM controllers, while augmenting them with serial link modeling and SMC-specific device logic to accurately represent the 3D-stacked memory organization and its embedded compute layer. The simulator provides a timing-accurate environment for modeling memory and interconnect subsystems, alongside functional modeling of the host SoC and in-memory processing cores. It supports scratchpad memories, DMA-driven data transfers, and complete address translation mechanisms (TLB/MMU), accompanied by a software stack comprising device drivers and user-level APIs that enable transparent workload offloading while maintaining virtual memory semantics.

At the architectural level, \texttt{SMCSim} partitions the SMC into multiple DRAM vaults, each equipped with local controllers and compute resources connected through a configurable logic-base interconnect. The framework integrates DRAM timing and bandwidth models, interconnect latency, and address-mapping policies, facilitating detailed exploration of system-level trade-offs such as data placement, thread scheduling, and inter-vault communication efficiency in 3D-stacked near-memory systems. Through its extensible interface, users can configure vault count, link bandwidth, and interconnect topology, making the simulator adaptable to a wide range of experimental scenarios.

Overall, \texttt{SMCSim} provides a reproducible, timing-aware platform for investigating Smart Memory Cube–style PIM architectures. It effectively captures host–memory interactions, data movement patterns, and interconnect behavior while maintaining a significantly lower computational overhead compared to full RTL models. Owing to this balance between accuracy and efficiency, \texttt{SMCSim} serves as a practical research tool for architectural exploration, performance analysis, and hardware/software co-design in emerging near-memory processing systems.

\subsection{PIMSim}
\texttt{PIMSim}~\cite{xu2018pimsim} is a comprehensive full-system simulation framework that supports a wide range of Processing-in-Memory (PIM) architectures. It addresses several limitations of earlier tools—such as fragmented toolchains and rigid hardware assumptions—by integrating multiple timing back-ends including \texttt{DRAMSim2}, \texttt{HMCSim}, and \texttt{NVMain}. Through this unified backend infrastructure, \texttt{PIMSim} enables co-simulation of hybrid memory systems such as DRAM, HMC, and non-volatile memories under a consistent timing and coherence model, thereby facilitating heterogeneous PIM system exploration within a single environment.

The simulator operates in three distinct modes that balance fidelity and performance according to design objectives. The first mode, known as \emph{Fast Mode}, targets rapid architectural prototyping. It employs pre-generated memory traces and simplified pipeline models to estimate performance and energy without modeling OS-level overheads, allowing researchers to test new logic designs or scheduling strategies within seconds. The second mode, \emph{Instrumentation-Driven Mode}, bridges trace-driven and full-system simulations. Leveraging Intel Pin as a front-end, it dynamically instruments running applications, identifies annotated PIM kernels at runtime, and redirects them to the simulator for online evaluation—offering realistic workload partitioning at moderate simulation cost. Finally, the \emph{Full-System Mode} extends \texttt{gem5} with PIM-specific instructions, detailed coherence protocols, and full system-call handling. It supports multiple ISAs (x86, ARM, SPARC), integrates fine-grained coherence mechanisms such as MESI and LazyPIM, and allows users to instantiate PIM logic as CPU-like cores, GPU units (via \texttt{GPGPU-Sim}), or abstract accelerators.

This hierarchical design enables researchers to transition seamlessly between fast design-space exploration and cycle-accurate full-system analysis within the same framework. By combining modular timing back-ends, flexible front-end interfaces, and broad architectural coverage, \texttt{PIMSim} provides a versatile and extensible foundation for investigating heterogeneous memory systems, host–PIM coherence, and near-data computation with consistent architectural semantics and reproducible results.

\subsection{UPMEM Functional Simulator }
\texttt{UPMEM Functional Simulator}~\cite{Oliveira2024UPMEMSDK} is an integral component of the official \texttt{UPMEM SDK}, developed to emulate the behavior of UPMEM’s Data Processing Units (DPUs) in the absence of physical hardware. The simulator is automatically activated when no DPU-equipped DIMM is detected in the system, though users can also enable it explicitly through the SDK environment configuration. It creates virtual ranks composed of one simulated chip per rank and supports up to 64 DPUs, maintaining compatibility with the same software stack used on real UPMEM hardware.

The simulator provides functional correctness rather than cycle accuracy. It executes compiled DPU binaries, emulating instruction behavior and host–DPU communication, but does not model timing, latency, or energy consumption. This design makes it highly suitable for application development, debugging, and verification of PIM programs prior to deployment on physical UPMEM modules. Since it faithfully reproduces the logical execution flow and data exchanges between the host and DPUs, the simulator operates across multiple abstraction levels, including application, system/algorithm, and architectural layers.

Overall, the \texttt{UPMEM Functional Simulator} offers a convenient and accessible software environment for developing, validating, and profiling near-memory computing workloads without requiring specialized hardware. It has become an essential tool for researchers and developers seeking to explore the UPMEM ecosystem and test large-scale memory-centric applications in a fully virtualized setup.

\subsection{Ramulator-PIM}
\texttt{Ramulator-PIM}~\cite{git2025ramulatorPIM} extends the widely adopted \texttt{Ramulator} DRAM simulator to incorporate cycle-accurate support for Processing-in-Memory (PIM) architectures. It retains Ramulator’s high-fidelity timing model while adding a dedicated flow that enables offloading of application kernels to simple in-order logic cores embedded within the 3D-stacked memory’s logic layer. The simulation operates in a trace-driven manner: a modified \texttt{ZSim} frontend executes the host program, producing two complementary traces—a filtered trace representing the CPU-side execution (capturing cache and coherence effects) and an unfiltered trace representing the PIM kernel execution. These traces are replayed within \texttt{Ramulator-PIM} to capture cycle-accurate memory timing, vault-level parallelism, and off-chip SerDes latencies. While the current public release does not support concurrent host–PIM execution, it serves as a precise framework for comparing kernel performance between host and PIM cores under realistic DRAM behavior.

To mitigate the inherently long runtime of full cycle-accurate simulations, \texttt{NAPEL} (Near-memory computing Application Performance \& Energy prediction via ensemble Learning)~\cite{Singh2019NAPEL} was introduced as a learning-based extension atop \texttt{Ramulator-PIM}. \texttt{NAPEL} leverages LLVM-based static analysis to extract a rich set of hardware-agnostic features from PIM kernels, such as instruction mix, memory reuse distance, and traffic patterns, and applies a Central Composite Design (CCD) methodology to select a minimal yet representative set of kernels and configurations for detailed simulation. The collected results are used to train a Random Forest model that accurately predicts performance (IPC) and energy for unseen kernels within the same near-memory configuration. Reported evaluations indicate prediction errors of approximately 8.5\% for performance and 11.6\% for energy, while achieving up to 220× faster design-space exploration compared to exhaustive simulation.

\texttt{Ramulator-PIM} employing \texttt{NAPEL} establishes a hybrid, calibration-driven workflow that combines the precision of cycle-accurate DRAM and logic-layer timing with the scalability of machine-learning-based performance prediction. This integration enables researchers to explore large PIM design spaces efficiently, maintaining analytical rigor without incurring the prohibitive cost of full-scale simulation across all workloads and configurations.

\subsection{MPU-Sim}
\texttt{MPU-Sim}~\cite{Xie2022MPUSim} is a cycle-accurate, research-grade simulator designed for evaluating near-bank processing in 3D-stacked DRAM architectures. It models a SIMT-style processor located on the base logic die and lightweight near-bank units (NBUs) integrated within individual DRAM dies. CUDA programs are compiled to PTX and executed on simple in-order subcores within the logic die, with selective offloading of computation to NBUs positioned adjacent to DRAM bank groups. This design exploits high internal bank bandwidth and minimizes data movement between compute and memory layers.

The simulator takes three main inputs, application program, data placement, and thread-block scheduling, to facilitate detailed studies on compiler back-end offloading, memory layout, and runtime scheduling strategies. Compared with prior open-source PIM frameworks, \texttt{MPU-Sim} incorporates near-bank specific features such as per-bank control logic, TSV arbitration, and a split execution pipeline for offloading, while targeting general-purpose SIMT workloads rather than fixed-function PIM kernels.

\texttt{MPU-Sim} generates fine-grained profiling traces and detailed component-level statistics that feed into static and dynamic energy models. Its components are cross-validated against well-established simulators: \texttt{GPGPU-Sim} for SIMT core behavior, \texttt{DRAMSim2} for DRAM timing, and \texttt{BookSim} for NoC interconnect modeling. Case studies highlight its analytical power—for example, demonstrating how DRAM refresh policies can significantly influence near-bank performance, and how aligning thread-block scheduling with bank mapping can yield substantial throughput gains.

Distributed as an open-source framework with containerized (Docker) setup scripts, \texttt{MPU-Sim} provides a robust and reproducible environment for exploring compiler, runtime, and architectural co-design in bank-level PIM systems. It stands as one of the few publicly available simulators combining CUDA/PTX programmability with detailed DRAM and NoC modeling, making it a valuable tool for studying the interaction between memory hierarchy organization and near-data compute efficiency.

\subsection{MultiPIM}
\texttt{MultiPIM}~\cite{Yu2021MultiPIM} is a detailed and highly configurable simulator for modeling large-scale Processing-in-Memory (PIM) systems built from multiple 3D-stacked memory devices. Unlike earlier single-stack frameworks such as \texttt{PIMSim} or \texttt{Ramulator-PIM}, it captures the realities of multi-stack interconnects, directory-based coherence, and virtual memory, all of which are essential to evaluate practical, scalable PIM designs.

Each memory stack is divided into several vaults, and every vault contains a lightweight in-order PIM core. Intra-stack connections, the local network that links vaults to the logic base of the same stack, are modeled as cycle-accurate crossbars and TSV links. Inter-stack connections, the high-speed links between different stacks, are configurable and simulated cycle-accurately using~\texttt{BookSim2}. The simulator’s backend, built on \texttt{Ramulator}, models DRAM timing and interconnect delays, while the frontend, based on \texttt{ZSim}, executes host instructions, manages thread scheduling, and drives memory traffic. This two-layer architecture gives \texttt{MultiPIM} both scalability and cycle-level accuracy for complex multi-stack memory systems. 

\texttt{MultiPIM} provides a source-level programming interface that allows developers to mark offloadable code regions with simple directives such as \texttt{pim\_blk\_begin/end()} or \texttt{pim\_mp\_begin/end()}. These APIs can automatically map POSIX/OpenMP threads or thread blocks onto the distributed in-memory PIM cores. The simulator supports user-defined interconnect topologies described via XML, and directory-based MESI coherence among PIM cores. It also integrates a unified virtual memory system, complete with TLBs and page table walkers, so both CPUs and PIM cores share a consistent address space. Profiling and statistics features help analyze latency, bandwidth bottlenecks, and coherence overheads in realistic applications.

\subsection{NeuroSim for PIM }
\texttt{NeuroSim}~\cite{Lu2021NeuroSim} is a functional, execution-driven, and modular simulation framework specifically developed for compute-in-memory (CIM) architectures employed in deep neural network (DNN) accelerators. It supports rapid and accurate design-space exploration across multiple levels of abstraction, from device physics to neural algorithm mapping, making it one of the most extensively adopted frameworks for evaluating resistive-memory-based neural accelerators.

Unlike single-layer simulators, \texttt{NeuroSim} integrates a hierarchical analytical modeling structure that spans four abstraction levels. At the \emph{device level}, it models memory cells and transistor technologies such as RRAM, PCM, and SRAM, incorporating non-ideal effects like variability and write asymmetry. At the \emph{circuit level}, it captures the behavior of peripheral modules, including sense amplifiers, ADC/DAC converters, and array-level interconnects. At the \emph{architecture level}, it organizes multiple subarrays into processing elements (PEs) and higher-order tiles interconnected through configurable routing networks. Finally, at the \emph{algorithm level}, it supports mapping and execution of arbitrary neural network layers for both inference and training, providing an end-to-end link between network behavior and underlying memory hardware.

This hierarchical approach enables \texttt{NeuroSim} to estimate key design metrics such as area, latency, energy consumption, and accuracy degradation under realistic device conditions. Validation against post-layout SPICE simulations of a 40\,nm RRAM-based CIM macro has demonstrated chip-level prediction errors below 1\% after calibration, confirming its reliability for pre-silicon evaluation.

An extended version, \texttt{DNN+NeuroSim}, further integrates the hardware model with PyTorch-based DNN training and inference frameworks, offering full-stack co-simulation from device to algorithm. This integration allows users to perform cross-technology comparisons and assess trade-offs between energy efficiency, latency, and accuracy across both digital and analog memory-based computing paradigms.

Overall, \texttt{NeuroSim} stands as a comprehensive functional and analytical simulator that bridges device-level modeling with system-level neural network performance analysis. Its modularity, validated accuracy, and broad scope make it a cornerstone tool for the study and optimization of CIM and PIM architectures in next-generation AI accelerators.

\subsection{PIMSimulator}
\texttt{PIMSimulator}~\cite{git2025PIMSimulator} is a cycle-accurate, \texttt{DRAMSim2}-backed memory simulator that models in-DRAM execution for HBM2-class systems. It augments a conventional DRAM timing substrate with a programmable SIMD PIM engine featuring scalar and vector register files (SRF/GRF) and a compact PIM ISA (e.g., \texttt{ADD}, \texttt{MUL}, \texttt{MAC}, \texttt{MAD}, \texttt{MOV}, \texttt{FILL}, \texttt{JUMP}). To coordinate near-data execution under realistic pseudo-channel timing, the simulator introduces PIM-specific DRAM commands (\texttt{Activate\_pim}, \texttt{ALU\_pim}, \texttt{READ\_pim}) that orchestrate data movement across banks, row buffers, and on-die compute units while preserving cycle-level fidelity.

By operating directly at the device/bank granularity, \texttt{PIMSimulator} enables quantitative studies of bank-level parallelism, in-memory data motion, and instruction-level kernel behavior without the confounding effects of full CPU/OS modeling. The resulting design space exploration remains lightweight yet timing-faithful for HBM-style PIM microarchitectures. In summary, \texttt{PIMSimulator} offers a concise, reproducible platform for prototyping and evaluating HBM-PIM kernels under realistic memory timing, complementing—but not replacing—full-system frameworks when OS, coherence, or multi-program effects are central to the research question.

\subsection{MemTorch }
\texttt{MemTorch}~\cite{Lammie2022MemTorch} is an open-source framework for simulating memristive (RRAM-based) deep learning systems, with a particular emphasis on modeling device-level non-idealities that affect neural network accuracy and reliability. Integrated natively with the PyTorch ecosystem, it enables users to train and evaluate conventional DNN and CNN models while transparently incorporating realistic hardware effects such as conductance drift, device variability, and limited precision into the computation flow.

\texttt{MemTorch} operates as a modular, execution-driven simulation environment. Instead of approximating abstract hardware performance, it executes PyTorch models directly through a simulation backend that intercepts layer operations at runtime to emulate in-memory multiply–accumulate (MAC) and convolutional operations within memristive crossbars. This allows the framework to capture the direct impact of device physics and circuit behavior on algorithmic outcomes during network execution and training.

The framework includes several physics-based device models—such as Linear Ion Drift, VTEAM, Stanford–PKU RRAM, and data-driven Verilog-A implementations—that describe conductance evolution, state transitions, and endurance degradation in resistive memory devices. By co-simulating both device-level non-idealities and peripheral circuit effects (including sensing, ADC/DAC quantization, and interconnect parasitics), \texttt{MemTorch} offers a highly realistic environment for studying how imperfections in emerging non-volatile memories influence the accuracy, robustness, and energy efficiency of neural networks.

\texttt{MemTorch} serves as a functional, modular, and execution-driven PIM-oriented simulator that spans multiple abstraction levels—from device and circuit modeling to algorithm and system integration. It provides researchers with a unified platform to investigate the effects of resistive memory variability and non-idealities on deep learning performance, enabling accurate hardware–software co-design and evaluation for next-generation compute-in-memory neural accelerators.

\subsection{PiMulator}
\texttt{PiMulator}~\cite{Mosanu2022PiMulator} is a modular and parameterizable FPGA-based platform built to accelerate the prototyping and evaluation of Processing-in-Memory (PIM) architectures with high fidelity. Implemented entirely in SystemVerilog and integrated within the LiteX SoC builder framework, \texttt{PiMulator} enables flexible configuration of both memory hierarchies and PIM compute logic. Researchers can instantiate custom architectures directly on FPGA hardware, adjusting timing, bandwidth, and interconnect parameters to reflect a wide range of memory technologies and design choices.

Operating across circuit- and architecture-level abstractions, \texttt{PiMulator} bridges low-level DRAM behavior with higher-level system evaluation. The framework models DDR4 and HBM memory subsystems, including channels, ranks, banks, and subarrays, while capturing timing-critical behaviors such as row activation, refresh, and inter-bank data movement. It further supports emulation of well-known DRAM-PIM mechanisms such as RowClone, Ambit, and LISA, facilitating direct comparison between compute- and data-movement-oriented approaches. With its event-driven hardware emulation pipeline, \texttt{PiMulator} achieves up to 28× runtime speedup compared to traditional software simulators, enabling large-scale, near-real-time exploration of complex PIM workloads.

Unlike software-only tools, \texttt{PiMulator} runs on FPGA fabric, providing cycle-faithful evaluation while maintaining configurability and modularity. Users can extend the base framework to include new compute primitives, interconnect topologies, or even non-volatile memory models through modular HDL components. This versatility makes it suitable for both architectural exploration and hardware/software co-design of PIM systems.

In summary, \texttt{PiMulator} is a functional, execution-driven, and modular hardware-accelerated simulation framework that offers researchers a bridge between simulation and silicon. By combining reconfigurability, speed, and fidelity, it serves as a practical open-source environment for evaluating DRAM- and HBM-based PIM designs, fostering rapid iteration from concept to prototype.

\subsection{MNSIM 2.0}
\texttt{MNSIM 2.0}~\cite{Zhu2023MNSIM2.0} is a behavior-level simulation framework developed to evaluate the performance, energy efficiency, and computational accuracy of Processing-in-Memory (PIM) architectures, particularly those used in AI acceleration. It addresses the growing need for rapid and accurate evaluation of memory-centric computing by eliminating the reliance on time-consuming circuit-level simulations. The tool provides a flexible, modular modeling environment that allows users to configure both digital and analog PIM arrays, peripheral circuits, and interconnect architectures, enabling comprehensive system-level performance assessment across large-scale neural workloads.

A major innovation of \texttt{MNSIM 2.0} lies in its unified array modeling structure, which supports heterogeneous memory technologies—such as RRAM-based analog crossbars and SRAM-based digital PIM units—under a single framework. This unified abstraction allows for consistent comparison between different compute-in-memory designs and facilitates hybrid evaluations where analog and digital subarrays coexist. The simulator integrates a PIM-oriented neural network training and quantization pipeline that co-optimizes neural models and hardware parameters, capturing device-level non-idealities, quantization noise, and limited precision effects during model development. Additionally, its scheduling interface enables flexible mapping of neural layers to hardware resources, supporting diverse dataflows (e.g., output-stationary, weight-stationary) and pipelined execution modes.

Operating at the behavioral level, \texttt{MNSIM 2.0} delivers results in seconds while maintaining high predictive accuracy, with reported deviation within 3.8\%–5.5\% compared to post-layout measurements of fabricated PIM macros. The framework provides detailed breakdowns of latency, power, energy, and accuracy degradation, helping designers identify performance bottlenecks and optimization opportunities. Its extensibility also allows integration with higher-level design frameworks and external deep-learning toolchains, making it suitable for end-to-end AI hardware–software co-design.

\texttt{MNSIM 2.0} is a functional, modular, and execution-driven simulation environment spanning architecture, system/algorithm, and application abstraction levels. It offers a fast yet accurate platform for evaluating PIM architectures in neural accelerators and supports comprehensive co-design exploration, bridging the gap between device behavior, architectural efficiency, and algorithmic robustness in next-generation AI computing systems.

\subsection{uPIMulator}
\label{sec:upimulator}
\texttt{uPIMulator}~\cite{Hyun2024uPIMulator} is a cycle-accurate simulator designed to execute real UPMEM binaries. It relies on the official UPMEM LLVM toolchain to compile C kernels, which are then run on a detailed model of the DPU (an in-DRAM core) which closely interacts with its DDR4 memory bank. The simulator models the 14-stage in-order pipeline, “revolver” thread scheduling, register hazards, and on-chip memories (WRAM, IRAM, MRAM). Host–device data transfers are characterized using asymmetric DMA bandwidths and realistic pseudo-channel timing, enabling researchers to investigate how kernels interact with both compute and memory resources under real hardware-like limitations.

\texttt{uPIMulator}~\cite{Hyun2024uPIMulator} is a cycle-accurate simulator calibrated for real UPMEM binaries. It uses the official UPMEM LLVM toolchain to compile C kernels and executes them on a detailed model of the DPU (an in-DRAM core) closely coupled to DDR4 memory banks, modeling a 14-stage in-order pipeline, “revolver” thread scheduling, register hazards, and on-chip memories (WRAM/IRAM/MRAM).
It executes actual DPU binaries and characterizes host–device transfers with asymmetric DMA bandwidth and realistic pseudo-channel timing, enabling analysis of compute–memory interactions under hardware-like constraints.

Beyond ensuring functional correctness, \texttt{uPIMulator} is designed for microarchitectural exploration. Its modular front-end and back-end allow researchers to prototype SIMT/vector execution, apply instruction-level parallelism (ILP) techniques such as superscalar issue and data forwarding, and experiment with memory system modifications like replacing scratchpads with caches. It also supports virtual memory through a lightweight MMU and TLB, enabling studies on multi-tenant isolation and pointer safety with only ~0.8\% performance overhead in published evaluations.
 
Overall, \texttt{uPIMulator} provides a practical, hardware-faithful platform for evaluating UPMEM-style PIM. It runs real ISA code on a configurable microarchitecture, making it ideal for analyzing compute–memory bottlenecks and testing architectural extensions such as vector units, ILP, virtual memory, and cache-based designs before hardware implementation.

\subsection{PIMeval}
\texttt{PIMeval}~\cite{10763591} is a lightweight, functional simulation framework designed to evaluate the performance and energy efficiency of digital DRAM-based Processing-in-Memory (PIM) architectures. Together with its companion benchmark suite, \texttt{PIMbench}, it provides a unified evaluation platform that allows fair and reproducible comparison across different PIM organizations. The framework supports multiple PIM design styles, including bit-serial subarray PIM, bit-parallel subarray (Fulcrum-style) PIM , and bank-level PIM. Through a unified PIM API, the toolchain abstracts low-level hardware differences, mapping high-level operations such as arithmetic, logical, and reduction functions into corresponding micro-operations specific to each PIM architecture. This abstraction layer enables seamless cross-platform evaluation, ensuring that applications written once can be executed and analyzed across multiple architectural variants.

Internally, \texttt{PIMeval} models performance and energy by dividing execution into two principal phases—data movement and computation. Timing analysis incorporates DRAM activation, precharge, and data-transfer latencies, while energy estimation combines standard DDR power models with additional contributions from on-die compute logic and interconnect circuitry. Although the current version relies on analytical models for latency estimation, future integration with \texttt{DRAMSim3} is planned to provide cycle-level timing accuracy. This makes \texttt{PIMeval} particularly suitable for early-stage design exploration and comparative analysis across digital PIM architectures.

The accompanying benchmark suite, \texttt{PIMbench}, provides a diverse set of workloads representative of key PIM application domains, including vector arithmetic, matrix operations, graph analytics, and machine learning kernels. Each workload is implemented once using the unified PIM API, ensuring portability and fair comparison across architectures. Together, \texttt{PIMeval} and \texttt{PIMbench} form a coherent ecosystem for PIM evaluation, bridging the gap between architectural modeling and application-level benchmarking, and enabling researchers to quantify performance and energy trade-offs in a consistent and reproducible manner.

\subsection{UPMEM LLM framework}
\texttt{UPMEM LLM framework}~\cite{ortega2024PIM-AI} is a functional simulation framework developed as part of UPMEM’s \texttt{PIM-AI} architecture, designed to accelerate large-scale language model (LLM) inference by integrating compute logic directly within DDR5 and LPDDR5 memory devices. Each \texttt{PIM-AI} chip can operate in two distinct modes: a \emph{Non-PIM mode}, functioning as a conventional 2\,GB DRAM device, and a \emph{PIM mode}, where it acts as a neural accelerator capable of executing core transformer operations within the memory subsystem itself. This dual-mode capability enables flexible deployment in both conventional memory hierarchies and heterogeneous AI accelerators.

The accompanying simulator provides a high-level functional modeling environment for evaluating LLM inference across various memory-centric architectures. It runs unmodified PyTorch models, enabling users to execute transformer-based workloads directly without conversion or retraining. The framework offers configurable hardware profiles representing different accelerator types (e.g., GPU, NPU, or custom PIM configurations) and collects performance statistics such as execution latency, data-transfer volume, energy consumption, and power. It models both the compute-bound encoding phase (typically dominated by GEMM operations) and the memory-bound decoding phase (dominated by GEMV operations), while also accounting for KV-cache movement and inter-layer dependencies. Additionally, the simulator supports flexible layer-to-hardware mapping, allowing hybrid execution across host and memory-resident compute elements.

Architecturally, the simulator adopts a modular, execution-driven design that captures the functional behavior of heterogeneous components without cycle-level timing or operating system emulation. This abstraction allows rapid design exploration, high-level profiling, and scalability studies of memory-centric inference systems. While not cycle-accurate, its high-level modeling fidelity provides valuable insight into performance and energy trade-offs in large-scale LLM deployments on PIM-enabled platforms.

\texttt{PIM-AI} offers researchers a practical environment for studying GPU-free inference acceleration through tightly integrated compute–memory architectures. By combining realistic workload execution with flexible configuration options, it serves as an effective prototyping and analysis tool for next-generation, memory-centric AI systems.

\subsection{NDPmulator}
\texttt{NDPmulator}~\cite{Vieira2024NDPmulator} is a gem5-based, cycle-accurate full-system simulator designed for detailed modeling and evaluation of Near-Data Accelerators (NDAcc) placed at various levels of the memory hierarchy—from caches to DRAM. It captures the complete CPU–memory–accelerator interaction, incorporating operating-system overheads, device-driver communication, and multi-process contention, thus enabling realistic performance and energy evaluation across diverse system configurations.

The simulator enhances \texttt{gem5} with two major extensions. The first is a \emph{Programming Interface (PI)} that abstracts CPU–accelerator coordination, managing synchronization, data transfers, and task invocation. The second is a \emph{Load/Store Unit (LSU)} that decomposes large accelerator memory transactions into smaller, fine-grained accesses, preserving coherence and timing fidelity while allowing NDAccs to be integrated seamlessly at any cache or memory level. Moreover, \texttt{NDPmulator} supports virtual memory translation and address remapping, ensuring transparent and consistent data sharing between CPU and accelerator domains.

The framework supports two complementary operation modes: (i) \emph{System Emulation (SE)}, which enables fast simulation of single-application workloads without OS overheads—ideal for early-stage microarchitectural exploration; and (ii) \emph{Full-System (FS)}, which runs a complete Linux stack, capturing driver latency, kernel scheduling, and resource contention for system-level accuracy. This dual-mode design allows researchers to trade off speed and fidelity depending on their study objectives.

Overall, \texttt{NDPmulator} provides a unified simulation infrastructure that bridges rapid design-space exploration with full-system realism, enabling comprehensive analysis of near-data accelerators and their interactions with modern memory hierarchies.

\subsection{M$^2$NDP}
\texttt{M$^2$NDP}~\cite{10764494} is a modular simulation framework designed to evaluate Near-Data Processing (NDP) architectures for Compute Express Link (CXL)–based memory systems. It introduces a Memory-Mapped NDP model that enables computation to occur directly within the CXL memory controller, thereby minimizing data movement and latency compared to conventional host-driven processing. The framework is built around two key architectural primitives: \texttt{M$^2$Func}, a \texttt{CXL.mem}-compatible communication interface between the host processor and the NDP controller, and \texttt{M$^2$µThread}, a lightweight microthreading engine designed to support fine-grained concurrency with minimal software overhead. Together, these mechanisms enable substantial performance and energy improvements—reportedly achieving up to 128× higher performance and 87.9\% lower energy consumption than CPU/GPU hosts equipped with passive CXL memory devices.

Constructed atop the \texttt{Ramulator} memory simulation core, \texttt{M$^2$NDP} integrates both functional and timing simulation capabilities, allowing users to model execution correctness while also capturing detailed latency, bandwidth, and power characteristics. The simulator adopts a modular design rather than a full-system one, making it particularly suitable for design-space exploration of individual PIM or NDP components, while remaining flexible for integration with higher-level architectural or system models. Its primary operational mode is trace-driven, enabling replay of real workload traces to evaluate memory-bound operations and kernel execution patterns within emerging CXL-attached architectures.

By combining accurate timing analysis with flexible, general-purpose NDP modeling, \texttt{M$^2$NDP} provides researchers with a practical and extensible platform for investigating the computational potential of memory-centric systems. Its modularity and adherence to CXL standards make it a valuable foundation for exploring next-generation heterogeneous systems, where memory devices play an increasingly active role in computation.

\subsection{PIMSIM-NN}
\texttt{PIMSIM-NN}~\cite{git2024PimSimNN} is a cycle-accurate and ISA-based simulation framework designed for evaluating PIM accelerators that execute deep neural network workloads. It addresses a key limitation of existing frameworks, where software and hardware models are tightly coupled, making it difficult to independently optimize compilation strategies or explore diverse hardware designs.

The central contribution of PIMSIM-NN is the introduction of a dedicated Instruction Set Architecture (ISA) that decouples software from hardware through a clean, abstract interface. Neural networks, described in ONNX format~\cite{bai2019}, are first compiled into ISA instructions by the accompanying PIMCOMP-NN compiler, which performs software-level optimizations and layer-to-hardware mapping. These ISA instructions are then executed by a cycle-accurate, event-driven simulator built on SystemC, which models configurable hardware parameters, realistic communication latency, and hardware parallelism through a reorder buffer mechanism.

This architecture enables flexible software–hardware co-design: researchers can test different compiler mappings without modifying the simulator, or sweep hardware configurations to assess energy, latency, and throughput trade-offs. Experimental validation shows that PIMSIM-NN captures realistic communication overheads (40–90\%), in contrast to prior simulators that underestimate this effect. By providing modularity, configurability, and open-source accessibility, PIMSIM-NN establishes a practical and extensible platform for the quantitative evaluation of PIM-based neural network accelerators.


\subsection{AiM Simulator}
\texttt{AiM Simulator}~\cite{Gu2025AiMsimulator} is a performance and power simulation framework specifically developed for Processing-in-Memory (PIM) and Processing-Near-Memory (PNM) architectures targeting large-scale artificial intelligence (AI) inference workloads, particularly large language models (LLMs). It serves as the foundation for the CENT (CXL-ENabled GPU-Free sysTem) project~\cite{Gu2025AiMsimulator}, which aims to replace conventional GPU-centric computation with memory-centric processing paradigms. The simulator models a hierarchical system of Compute Express Link (CXL)–attached memory modules, where each module integrates near-bank PIM units optimized for dense linear operations (e.g., multiply–accumulate) and PNM units composed of lightweight RISC-V cores for non-linear functions such as normalization and Softmax. Through support for key CXL communication primitives (send, receive, and broadcast), \texttt{AiM Simulator} captures both intra-module and inter-module data movement, enabling realistic evaluation of distributed inference across large-scale multi-stack memory systems.

Architecturally, \texttt{AiM Simulator} performs cycle-level modeling of DRAM timing, PIM/PNM computation, and CXL link delays, while maintaining scalability across multiple modules. It also incorporates bandwidth and coherence constraints intrinsic to the CXL protocol, allowing researchers to investigate host–device communication overheads during neural layer offloading. Moreover, the simulator facilitates exploration of various parallelization strategies—pipeline, tensor, and hybrid parallelism—adopted in CENT-style systems. Integrated power and cost models allow users to estimate latency, throughput, and energy consumption under different hardware configurations, including link bandwidth, number of modules, and compute density.

Experimental evaluations from associated studies demonstrate that the CENT architecture, modeled using \texttt{AiM Simulator}, achieves more than 2× higher inference throughput and approximately 3× better energy efficiency than NVIDIA A100 GPUs for large-scale LLM inference, while also revealing critical performance bottlenecks imposed by CXL bandwidth and latency. By combining detailed timing models for PIM/PNM devices with accurate interconnect and power modeling, \texttt{AiM Simulator} provides a versatile and practical platform for designing, analyzing, and optimizing GPU-free, memory-centric architectures for next-generation large-scale AI systems.

\subsection{3D CiM LLM Sim}
\texttt{3D CiM LLM Sim}~\cite{Buchel2025EfficientScaling} is a high-level, functional performance simulator designed to evaluate large language model (LLM) inference—including both dense and Mixture-of-Experts (MoE) architectures—on a conceptual 3D Analog In-Memory Computing (AIMC) accelerator. Its objective is to characterize latency, energy efficiency, and memory utilization when executing transformer-style networks on vertically stacked, high-density non-volatile memory substrates.

The simulator operates through a structured four-stage pipeline encompassing model definition and mapping, graph tracing, graph processing, and execution scheduling. Initially, the LLM model is defined in PyTorch and mapped onto analog compute tiles and digital processing units using a greedy utilization strategy that maximizes hardware efficiency. The system then leverages \texttt{torch.fx} to extract a fine-grained computational graph. During graph processing, high-level matrix–vector multiplication (MVM) operations are decomposed into tier-level sub-operations to accurately represent the 3D AIMC hierarchy. In the final scheduling phase, a custom instruction scheduler enforces architectural constraints such as one-tier-at-a-time (OTT) execution and bounded DPU or multi-head attention (MHA) availability, producing cycle- and resource-aware execution timelines.

While \texttt{3D CiM LLM Sim} models computational and dataflow behavior in detail, it intentionally abstracts interconnect contention and assumes a constant per-bit transfer cost between compute and memory tiers to preserve architecture generality. This design choice allows for flexible adaptation to multiple 3D AIMC configurations while avoiding overfitting to a specific interconnect implementation.

\texttt{3D CiM LLM Sim} offers a flexible, modular, and execution-driven simulation environment that enables early-stage architectural exploration of analog and hybrid 3D in-memory accelerators. It serves as a practical tool for analyzing LLM inference performance across different AIMC hierarchies, providing key insights into latency–energy trade-offs, memory access patterns, and utilization efficiency in next-generation compute-in-memory designs.



\section{Simulation Perspectives: Functional vs. Timing Simulators}
\label{sec:sim_perspectives}

Understanding the distinction between functional and timing simulators is fundamental to accurate modeling and verification of modern digital and memory-centric systems. These two perspectives are inherently complementary: functional simulators ensure the correctness of logical behavior, whereas timing simulators validate performance and reliability under realistic microarchitectural constraints. As system complexity grows and timing margins shrink, leveraging both perspectives becomes essential for robust and efficient PIM design and validation. The classification of PIM simulators into functional and timing-oriented categories, along with representative examples, is presented in Table~\ref{tab:Functional_Timing}, while the subsequent sections provide an in-depth discussion of each simulation perspective. 

\begin{table}[h]
\centering
\caption{Functional and Timing PIM Simulators}
\label{tab:Functional_Timing}
\resizebox{\textwidth}{!}{
\begin{tabular}{|
    >{\centering\arraybackslash}m{5cm}| 
    >{\arraybackslash}m{14cm}| 
}
\hline
\textbf{Simulator Perspectives} & \multicolumn{1}{c|}{\textbf{Representative Simulators}} \\ 
\hline
\textbf{Functional} &
PIMeval, PIMSim (Fast), NeuroSim, MemTorch, UPMEM SDK functional Sim, 3D CiM LLM Sim, UPMEM LLM Framework, MNSIM 2.0 \\
\hline
\textbf{Time-based} &
NDPmulator, PIMSim (Full, Inst), M$^2$NDP, PIMSimulator, uPIMulator, MPU-Sim, SMCSim, NVmain, PIMSIM-NN, MultiPIM, Ramulator-PIM, PiMulator, AiMsimulator \\
\hline
\end{tabular}
}
\end{table}

\subsection{Functional Simulators}
Functional simulators model the logical behavior of an instruction set architecture (ISA) or computational model without considering the temporal evolution of events. Instructions are executed in program order, ensuring architectural correctness—inputs and outputs match those of the real hardware—but without modeling microarchitectural timing effects such as pipelining, cache latency, or DRAM access delay.

Since functional simulators omit timing behaviour, they are lightweight and fast, making them particularly useful in the early stages of PIM designs. Researchers rely on them to validate the correctness of a new ISA extension, test compiler or runtime modifications, and ensure that software stacks execute properly on top of PIM instructions \cite{Buitrago_Com_2024,8718630}. Virtual platforms and early MPSoC simulators are typical examples of this class of simulators.

The trade-off is that functional simulators cannot capture performance-critical phenomena such as cache contention, memory bank conflicts, or the overlap between host CPU execution and in-memory computation. They also often limit themselves to user-level execution, skipping OS-level behavior such as interrupts or system calls \cite{10.5555/1855084}.

Although functional simulators do not capture detailed cycle-level timing, many modern PIM-oriented frameworks integrate analytical delay and latency estimators to provide approximate performance insight without the complexity of full microarchitectural modeling. 
Rather than advancing time cycle by cycle, tools such as \texttt{NeuroSim}, \texttt{MemTorch}, \texttt{PIMeval} and the Fast Mode of  \texttt{PIMSim} utilize closed-form analytical equations or calibrated look-up tables based on \texttt{SPICE}~\cite{Nagel1973SPICE} or \texttt{CACTI}~\cite{Balasubramonian2017CACTI} data to estimate access delay, energy, and throughput for memory-array operations~\cite{git2024NeuroSim, git2022MemTorch, git2025PIMevalPIMbench, git2022PIMSim}.

At this abstraction level, latency is computed from analytical models that relates circuit and architectural parameters such as cell resistance, wordline and bitline capacitance, and data reuse ratio, rather than through detailed cycle-by-cycle timing updates.

This hybrid analytical approach effectively bridges the gap between purely functional validation and comprehensive timing simulation, enabling functional tools to maintain high simulation speed while delivering quantitatively meaningful performance trends for efficient design-space exploration.

In summary, functional simulators are primarily designed to verify the correctness of system behavior, providing a fast and efficient means to validate functional accuracy. However, they lack the detailed timing information necessary to accurately assess performance metrics and provide the critical insights required for comprehensive performance analysis and architectural optimization. In contrast, timing simulators incorporate cycle-level timing details, enabling precise evaluation of performance characteristics such as latency, throughput, and energy efficiency. This distinction makes timing simulators essential for quantifying how much performance PIM architectures can deliver relative to traditional CPU and GPU baselines. 

\subsection{Timing Simulators}
Timing simulators extend beyond functionality by modeling how instructions progress through microarchitectural structures such as pipelines, memory hierarchies, interconnects, and PIM units. Often called performance simulators, they provide detailed metrics such as latency, throughput, and energy efficiency. In this era, two main approaches exist: a) cycle-level and b) event-driven simulations, which will be discussed in detail below.

\subsubsection{Cycle-level simulation}
This kind of simulation advances the system one cycle at a time, updating every modeled component. It offers high fidelity by capturing pipeline hazards, memory scheduling, and contention making it particularly useful for studying fine-grained CPU–PIM interactions.
However, cycle-level models are computationally expensive and memory-intensive, limiting their practical use to shorter workloads or micro-benchmarks \cite{7753351,8718630}.

Cycle-based timing simulators usually achieve cycle accuracy by employing event scheduling and timing progression governed by validated memory timing backends that model the detailed behavior of modern DRAM and NVM subsystems.
In practice, many PIM and near-memory simulation frameworks rely on well-established backends such as \texttt{Ramulator}~\cite{Kim2015Ramulator}, \texttt{DRAMSim2/3}~ \cite{Rosenfeld2011DRAMSim2,Li2020DRAMsim3}, \texttt{HMCSim}~\cite{Leidel2016HMCSim}, \texttt{NVMain}~\cite{Poremba2012NVMain} and \texttt{BookSim}~\cite{Jiang2013booksim} to maintain cycle-accurate synchronization across compute, interconnect, and memory domains. Each backend incorporates a dedicated timing engine:
\begin{itemize}
    \item \texttt{Ramulator} models DRAM commands (ACT, PRE, RD, WR) and enforces JEDEC timing parameters at sub-bank granularity, enabling microsecond-scale trace replay with deterministic per-cycle scheduling.
    \item \texttt{DRAMSim2} introduced a transaction-level model where every DRAM command updates internal counters each clock cycle. \texttt{DRAMSim3} extends this to multi-channel, multi-rank DDR4/LPDDR4 systems with cycle-accurate queue arbitration and detailed power accounting.
    \item \texttt{HMCSim} abstracts 3D-stacked Hybrid Memory Cube interfaces, capturing vault controllers, link serialization, and TSV latencies at cycle granularity, enabling realistic timing analysis for near-bank PIM logic.
    \item \texttt{NVMain} targets non-volatile technologies (PCM, STT-RAM, ReRAM), combining cycle-accurate command scheduling with device-specific write/ erase delays and endurance models.
    \item \texttt{BookSim} complements these by modeling network-on-chip (NoC) or interconnect fabric at a cycle level, synchronizing packet injection, router arbitration, and flit propagation with the memory backend.
\end{itemize}

As an example, \texttt{MultiPIM} and \texttt{Ramulator-PIM} leverage  \texttt{Ramulator} to reproduce precise DRAM command scheduling and vault-level contention~\cite{Yu2021MultiPIM, git2025ramulatorPIM}. \texttt{PiMulator} implements a DRAM controller RTL on FPGA, calibrates its parameters against \texttt{DRAMSim3} for clock-synchronous operation~\cite{Mosanu2022PiMulator}, while \texttt{MPU-Sim} adopts \texttt{BookSim} for network timing and couples it with DRAMSim2 for memory access latency modeling~\cite{Xie2022MPUSim}.

Conversely, some simulators achieve cycle accuracy without relying on existing backend.
For instance, \texttt{uPIMulator} employs a fully custom event scheduler that models every DPU pipeline stage and DDR transaction explicitly in its timing engine, reaching hardware-validated precision through direct comparison with real UPMEM modules~\cite{Hyun2024uPIMulator}.

Such cycle-accurate architectures are essential for evaluating fine-grained interactions between compute pipelines, memory controllers, and in-memory execution units, providing deterministic timing behavior essential for hardware–software co-designs verification.

Hybrid approaches often combine cycle-level memory modeling with functional representations of computational circuits to balance accuracy and simulation efficiency \cite{7753351}. In this work, we adopt cycle-accurate models consistent with these hybrid methodologies.

\subsubsection{Event-driven simulation}
This kind of simulation models advance simulation time by skipping idle cycles and progressing only on relevant events, e.g., memory requests or PIM kernel calls. This significantly improves simulation speed and scalability, albeit with reduced cycle-level detail making it particularly well-suited for large workloads and application-scale PIM evaluations \cite{Buitrago_Com_2024}.

Unlike cycle-accurate simulators that update system state every clock cycle, event-driven simulators manage discrete events—such as command completions, kernel activations, or DMA transfers—and focus computational resources solely on operations that impact architectural state.
This paradigm has gained prominence in large-scale and heterogeneous PIM simulators, where full cycle-level modeling becomes computationally prohibitive.

Among contemporary PIM simulators, only a subset implements true event-driven kernels.
For example, \texttt{LLMServingSim}~\cite{10763697} explicitly builds upon the \texttt{ASTRA-Sim 2.0}~\cite{won2023astra} infrastructure, which utilizes a discrete-event engine to coordinate interactions among GPUs, NPUs, and PIM with timing granularity ranging from milliseconds to nanoseconds. Simulation time in this framework advances through  event queues such as ``kernel start,'' ``DMA completion,'' and ``token generation finish'' rather than cycle-by-cycle, rendering it event-driven with cycle-level granularity but not strictly cycle-accurate.

\texttt{LLMServingSim} extends its event-driven co-simulation to heterogeneous architectures that integrate PIM modules for accelerating memory-bound GEMV operations in LLM attention layers. This feature enables evaluating NPU–PIM cooperative execution within large-scale LLM serving environments~\cite{10763697}.

\texttt{PIMulator-NN}~\cite{zheng2022pimulator} introduces its own event queue core (EDSC), where every computation and data-transfer module generates ``START,'' ``STOP'' and ``FINISH'' events. Latency and energy consumptions for these events are estimated using backend analytical models (e.g. \texttt{NeuroSim}, \texttt{CACTI}, and \texttt{NVSim}).
This makes PIMulator-NN a canonical event-driven and cross-level simulator that integrates circuit-, architecture-, and algorithm-level timing without continuous clock stepping.
However, since the source codes of these simulators are not publicly available, they are excluded from detailed examination in the subsequent chapters.

In the context of PIM research, timing simulators are indispensable for quantifying performance benefits such as the number of cycles saved by near-memory execution and the impact on bandwidth and energy efficiency. Consequently, event-driven and timing-accurate models are critical for rigorous performance evaluation of PIM architectures.

\section{Simulation Scope: Full-System vs. Modular Simulation}
\label{sec:sim_scope}

Simulators can also be classified according to their scope and target, distinguishing those designed to model the entire hardware/software stack (full-system simulators) from those focusing on specific subsystems or user-level applications (modular simulators), as shown in Table~\ref{tab:pim-scope}. This distinction critically influences their applicability in PIM research, determining whether a tool is suited for comprehensive system-level evaluation or for isolated studies of memory-compute interactions.

\begin{table}[h]
\centering
\caption{Full-System and Modular PIM Simulators}
\label{tab:pim-scope}
\resizebox{\textwidth}{!}{
\begin{tabular}{|
    >{\centering\arraybackslash}m{4cm}|
    >{\arraybackslash}m{14cm}|
}
\hline
\textbf{Simulator Scope} & \multicolumn{1}{c|}{\textbf{Representative Simulators}} \\ 
\hline
\textbf{Full-System} &
PIMSim, NDPmulator, SMCSim, uPIMulator, MultiPIM, M$^2$NDP \\ 
\hline
\textbf{Modular} &
PIMSim, PIMeval, PIMSimulator, MPU-Sim, PiMulator, AiM Simulator, PIMSIM-NN, NeuroSim, MemTorch, NVMain, UPMEM SDK functional Sim, UPMEM LLM Framework, 3D CiM LLM Sim, MNSIM 2.0, Ramulator-PIM \\
\hline
\end{tabular}
}
\end{table}

\subsection{Full-System Simulators}
Full-system simulators emulate an entire computing platform, encompassing processors, memory subsystems, I/O devices, and the operating system.
Such simulators are cable of booting an operating system within the simulation environment, allowing the execution of a complete software stack from kernel services to user applications, thus replicating real hardware behavior \cite{8718630,   10.5555/1855084}.

This comprehensive modeling enables the investigation of OS-level effects—including system calls, virtual memory management, scheduling policies, and device drivers—which are particularly relevant for PIM architectures where near-memory computation integrates tightly with system software.
A prominent example is gem5, which supports both full-system and syscall emulation modes. 

In full-system mode, gem5 provides detailed analysis of PIM performance under realistic OS scheduling and memory management, whereas syscall emulation offers a lighter-weight environment for user-level studies \cite{8718630}. Similarly, SimOS delivers a functional full-system simulation environment \cite{10.1145/244804.244807}.

The principal limitation of full-system simulation lies in its complexity and computational overhead. Booting and managing an OS with all associated devices significantly increases simulation run-time, often confining full-system simulation to detailed case studies rather than rapid prototyping \cite{7753351}. Nevertheless, when accuracy and realism are essential, such as for evaluating PIM accelerators integration with heterogeneous systems, full-system simulation is indispensable.

\subsection{Modular Simulators}
In contrast, modular simulators restrict their focus to specific modules or user-level applications, typically bypassing the operating system by intercepting system calls on the host machine \cite{8718630,Buitrago_Com_2024}.
This narrower scope substantially reduces simulation complexity and accelerates execution.

Modular simulators are particularly valuable for targeted PIM investigations centered on memory subsystem behavior or in-memory computation, without the need to simulate full OS interactions.
For instance, ZSim is a fast timing simulator for the x86 ISA that operates at the application level. Several PIM-oriented frameworks,including \texttt{MultiPIM} \cite{Yu2021MultiPIM} and \texttt{NATSA} \cite{Fernandez2020NATSA}, extend \texttt{ZSim} with \texttt{Ramulator} \cite{git2025ramulatorPIM} to model near-memory computation, focusing on memory subsystem dynamics and PIM kernel execution.

The trade-off for modular simulators is reduced accuracy, as they omit OS-level effects such as scheduling overheads, memory-mapped I/O, and interrupt handling, which can significantly impact server and data-intensive workloads \cite{7753351}. As a result, modular simulation is well-suited for rapid design space exploration of PIM primitives or microarchitectural components but inappropriate for assessing full-system deployment scenarios.

\subsection{Case Study: PIMSim (full-system and modular in one framework)}

PIMSim exemplifies a flexible simulator that integrates detailed PIM logic, memory devices, and coherence into a unified architecture. It combines DRAMSim2~\cite{Rosenfeld2011DRAMSim2}, HMCSim~\cite{Leidel2016HMCSim}, and NVMain~\cite{git2018NVMain} backends, incorporates PIM-enabled instructions and code annotations, and supports both fine-grained (MESI) and coarse-grained coherence models. PIMSim provides three operational modes, each balancing fidelity and speed:

\begin{itemize}
    \item Full-System mode: Executes user and kernel code by booting an OS inside gem5, modeling processors, memory, peripherals, and PIM logic in detail. It implements PIM pseudo-instructions and executes detailed PIM-enabled operations with coherence protocols (MESI, page, or LazyPIM-like\cite{7485993}). This mode is essential when OS effects such as virtual memory, scheduling, and drivers critically influence PIM behavior.
          
    \item Instrumentation-driven (Pin-based) mode: Rather than simulating a full OS, PIMSim employs Intel Pin to monitor a running program on a host machine. It records dynamic instruction and memory behaviors as they happen and feeds them directly into the simulator, capturing realistic runtime dynamics with reduced overhead compared to full-system simulation.
 \item Fast (trace-driven) mode: 
 This mode bypasses program execution entirely, instead replaying pre-recorded trace files containing logged memory accesses and instructions. This approach enables rapid evaluation of multiple hardware configurations but sacrifices the ability to model dynamic runtime effects.
\end{itemize}


\section{Workload Feeding: Execution-driven vs. Trace-driven }
\label{sec:sim_workload}

\begin{table}[h]
\centering
\caption{Execution- and Trace-Driven PIM Simulator}
\label{tab:pim-workload}
\resizebox{\textwidth}{!}{
\begin{tabular}{|
    >{\centering\arraybackslash}m{5cm}|
    >{\arraybackslash}m{14cm}|
}
\hline
\textbf{Workload Feeding Type} & \multicolumn{1}{c|}{\textbf{Representative Simulators}} \\ 
\hline
\textbf{Execution-Driven} &
PIMeval, PIMSim (Full, Inst), NDPmulator, MNSIM 2.0, M$^2$NDP, PIMSimulator, uPIMulator, MPU-Sim, MemTorch, SMCSim, PIMSIM-NN, MultiPIM, NeuroSim, UPMEM SDK functional Sim, UPMEM LLM Framework \\
\hline
\textbf{Trace-Driven} &
PiMulator, PIMSim (Fast), AiM Simulator, NVMain, 3D CiM LLM Sim, Ramulator-PIM \\
\hline
\end{tabular}
}
\end{table}

Simulators can be categorized based on their input source distinguishing those that execute binaries directly (execution-driven) from those that replay pre-recorded instruction and memory traces (trace-driven).
This classification has a significantly impact on simulator speed, accuracy, and flexibility, particularly in the context of PIM systems, where workloads often involve large-scale memory operations.\\

\subsection{Execution-driven Simulators}
Execution-driven simulators dynamically run benchmark binaries on the simulated architecture, integrating functional correctness with timing models.
By directly executing programs, these simulators avoid reliance on pre-generated traces, simultaneously tracking performance metrics such as instruction latencies, memory accesses, and PIM kernel execution times \cite{7753351}\cite{10.5555/1855084}.
This approach offers several advantages. It eliminates the overhead of generating and storing massive trace files, which is particularly beneficial for data-intensive PIM operations \cite{8718630}. Moreover, execution-driven simulators can naturally capture behaviors dependent on runtime conditions, including speculative execution, dynamic scheduling, and branch mispredictions, which static trace-driven models cannot fully represent.

However, the complexity of maintaining a combined functional and timing engine makes execution-driven simulators typically slower and more challenging to develop \cite{Buitrago_Com_2024}.
Despite this, they are invaluable for analyzing runtime interactions between memory-side PIM operations and host processors, as well as for studying fine-grained effects such as coherence, consistency, and synchronization.

Several execution-driven frameworks have been proposed within the PIM domain.
For instance, \texttt{PiMulator} executes workloads directly on a cycle-accurate FPGA-based SoC and models PIM timing alongside host processor interactions, enabling comprehensive studies of system-level phenomena such as coherence and runtime scheduling~\cite{Mosanu2022PiMulator}.
Similarly, \texttt{PIMSim} offers a full-system execution-driven mode that supports both user- and kernel- level code to simulate complete PIM-aware systems~\cite{xu2018pimsim}.

The \texttt{UPMEM LLM framework} integrates with PyTorch to overrides layer execution at runtime, directly executing models such as LLaMA or Mixtral, while tracking compute, bandwidth and energy on a per-layer basis~\cite{git2024UPMEM_LLM_Framework}. 
Likewise, \texttt{M$^2$NDP} runs real RISC-V kernels and full PIM workloads cycle by cycle, modeling both functional correctness and detailed timing in CXL-attached near-data processors~\cite{10764494}.
These tools demonstrate how execution-driven simulations enables observation of dynamic runtime effects, including synchronization, contention, and data orchestration between host and memory-side compute units, which are not accessible through static trace replay.

\subsection{Trace-driven Simulators}
\begin{figure}[h]
\centering
\includegraphics[scale=0.35]{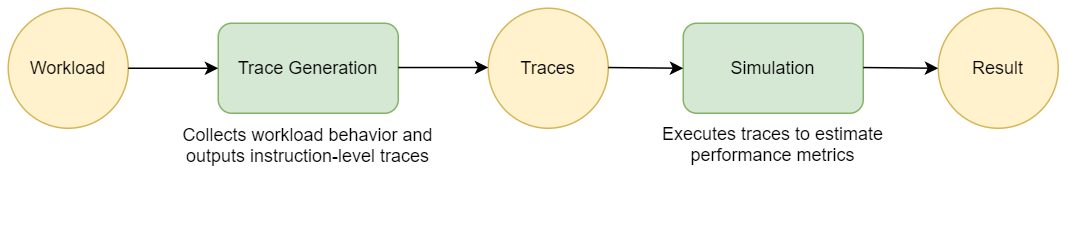}
\caption{Abstract view of trace-driven simulation.}
\label{fig:trace-driven}
\end{figure}

Trace-driven simulators rely on pre-recorded instruction or memory-access traces generated from benchmark runs on real hardware or functional simulators \cite{8718630}. These traces, comprising sequences of instructions, data addresses, and control flow outcomes, are input into a timing model to estimate performance (See Figure~\ref{fig:trace-driven}).
This separation of functional and timing simulation offers a key advantage: once generated, traces can be reused to evaluate numerous architectural configurations without re-executing the program \cite{Buitrago_Com_2024}.
This reusability accelerates design-space exploration, which is particularly advantageous in PIM research where a single memory-access trace can be replayed across multiple timing models to compare near-memory computation schemes.

Nevertheless, trace-driven simulators encounter several limitations.Trace files can be extremely large and require significant storage and processing resources, sometimes necessitating sampling or compression techniques \cite{7753351}. More critically, trace-based methods cannot runtime-dependent behaviors such as speculative execution or dynamic resource contention \cite{10.5555/1855084}.
Consequently, while trace-driven approaches effectively characterize bulk memory access patterns and support comparative evaluations of PIM array designs, they may overlook crucial timing interactions between PIM modules and host CPUs.

Trace-driven methods are prevalent in PIM simulation due to their scalability. Tools such as \texttt{PIMSim} (fast mode), \texttt{Ramulator-PIM}, \texttt{NVMain}, and \texttt{AiMSimulator} rely on pre-collected traces to decouple functional and timing simulations. This enables rapid iteration over different memory organizations, near-memory instruction sets, and device-level latency models without costly benchmark re-execution~\cite{xu2018pimsim, git2025ramulatorPIM, Poremba2012NVMain, Gu2025AiMsimulator}. Additionally, trace-driven simulators are well-suited to large-scale AI workloads; for example, \texttt{AiMSimulator} and IBM’s 3D CiM LLM simulator generate instruction and memory traces from transformer-based LLMs, replaying these to evaluate PIM hardware performance in end-to-end inference tasks~\cite{Buchel2025EfficientScaling}. Similarly, \texttt{NVMain} supports non-volatile memory research by feeding in CPU-generated traces from simulators like gem5~\cite{Poremba2012NVMain}.
These examples highlight how trace-driven PIM simulators efficiently capture high-level dataflow patterns and energy–latency trade-offs, albeit at the expense of some fidelity in modeling runtime-dependent behaviors.

\section{Abstraction Levels in PIM Simulation}
\label{sec:sim_abstractions}

Simulation of PIM systems encompasses multiple abstraction levels, each providing distinct insights and trade-offs between accuracy, complexity, and runtime. Understanding these abstraction layers is essential for selecting the appropriate simulation methodology depending on the research goals—ranging from high-level algorithmic performance analysis to detailed circuit- and device-level evaluations. We classify PIM simulation abstraction levels into five primary categories: Application level, System/Algorithm level, Architecture level, Circuit level, and Device level.
Each of these abstraction levels is discussed below, and representative simulators are presented in Table~\ref{tab:pim-abstraction}.

\begin{table}[h]
\centering
\caption{Abstraction Levels in PIM Simulation}
\label{tab:pim-abstraction}
\resizebox{\textwidth}{!}{
\begin{tabular}{|
    >{\centering\arraybackslash}m{5cm}|
    >{\arraybackslash}m{14cm}|
}
\hline
\textbf{Abstraction Level} & \multicolumn{1}{c|}{\textbf{Representative Simulators}} \\
\hline
\textbf{Application} &
AiM Simulator, MNSIM 2.0, NeuroSim, MemTorch, Ramulator-PIM, UPMEM LLM Framework, 3D CiM LLM Sim \\
\hline
\textbf{System / Algorithm} &
AiM Simulator, MNSIM 2.0, MultiPIM, NeuroSim, PiMulator, Ramulator-PIM, MemTorch, MPU-Sim, M$^2$NDP, NDPmulator, PIMeval, PIMSim, SMC-Sim, uPIMulator, UPMEM SDK functional Sim, PIMSimulator, UPMEM LLM Framework, 3D CiM LLM Sim \\
\hline
\textbf{Architecture} &
AiM Simulator, MNSIM 2.0, MultiPIM, NeuroSim, NVMain, PiMulator, MemTorch, MPU-Sim, M$^2$NDP, NDPmulator, PIMeval, PIMSim, SMC-Sim, uPIMulator, UPMEM SDK functional Sim, PIMSimulator, 3D CiM LLM Sim \\
\hline
\textbf{Circuit} &
AiM Simulator, MNSIM 2.0, NeuroSim, NVMain, PiMulator, MemTorch, MPU-Sim, M$^2$NDP \\
\hline
\textbf{Device} &
MNSIM 2.0, NeuroSim, NVMain, MemTorch \\
\hline
\end{tabular}
}
\end{table}

\subsection{Application Level} 
At the highest abstraction, application-level simulation models the behavior of end-user programs or workloads running on PIM-enabled systems.
This level focuses on capturing application characteristics such as data flow, memory access patterns, and computational kernels without detailed consideration of hardware timing or architectural constraints.
Application level evaluates end-user workloads such as neural networks, graph analytics, or Boolean processing.

In another word, this level's simulators typically integrate with machine learning frameworks or runtime environments to analyze how PIM affects overall application performance and energy consumption.
It also provides application-level metrics (throughput, energy savings) on top of predefined PIM architectures (e.g., MNSIM, NeuroSim) \cite{maurer2025survey}.
This abstraction is particularly useful for early-stage exploration of workload suitability for PIM acceleration and for guiding architectural design choices based on observed algorithmic behavior \cite{Chandrasekaran2017}.

\subsection{System/Algorithm Level}
System or algorithm-level simulation abstracts the PIM architecture to evaluate the impact of PIM operations within the context of the broader system or algorithm.
This level models key components such as host processors, memory subsystems, and interconnects while incorporating simplified PIM kernels or instruction sets. System-level simulators capture interactions between PIM modules and the CPU, including data movement, scheduling, and resource contention, enabling the study of system-wide effects such as bandwidth utilization and latency reduction.
\cite{Bertolli2016}.
At this level of abstraction, simulators demonstrate how operations or kernels (e.g., DNN layers, PIM instructions) are mapped and scheduled onto PIM hardware (e.g., PIMSim, PIMulator-NN) \cite{maurer2025survey}.
This level balances simulation speed with the ability to analyze performance and scalability of various algorithms when offloaded to memory-side compute units.

\subsection{Architecture Level}
Architecture-level simulation provides detailed modeling of the PIM microarchitecture, including components such as memory controllers, compute units, interconnect fabrics, and coherence protocols.
This abstraction emphasizes cycle-level timing, pipeline behavior, and memory hierarchy interactions.
It describes how PEs are organized into banks, channels, and interconnects. Explores system-wide design trade-offs such as scalability, bandwidth, and hierarchy, e.g., NVSim \cite{maurer2025survey}.
Architectural simulators enable precise evaluation of design trade-offs such as instruction set extensions, cache coherence policies, and near-memory processing logic \cite{Park2019}. They are essential for verifying the functional correctness and performance of proposed PIM designs and for guiding hardware implementation decisions.

\subsection{Circuit Level}
Circuit-level simulation models PIM at the granularity of logic gates, timing delays, and electrical characteristics of integrated circuits.
This level involves detailed analysis of critical circuit blocks such as sense amplifiers, arithmetic logic units, and memory cell arrays, accounting for signal propagation delays, switching activities, and power consumption.
Circuit simulators often use tools like SPICE to validate the timing and energy efficiency of proposed PIM circuits  \cite{Li2020}. This abstraction is crucial for optimizing circuit designs, minimizing latency and energy, and ensuring reliable operation under process variation and environmental conditions \cite{maurer2025survey}.

\subsection{Device Level}
At the lowest abstraction, device-level simulation focuses on the physical and material properties of the memory technologies used in PIM, such as DRAM, SRAM, PCM, ReRAM, or STT-RAM \cite{dorostkar2024empirical}.
This level models phenomena including charge storage, retention, resistance switching, and endurance under electrical stress  \cite{maurer2025survey}. Device simulators incorporate physics-based models to predict device behavior, variability, and failure modes, which directly influence circuit and architectural design choices. Accurate device-level models are indispensable for emerging memory technologies and for enabling reliable integration of novel memory devices in PIM architectures \cite{Zhang2021}.

This hierarchical framework highlights the complementary nature of abstraction levels in PIM simulation, enabling researchers to choose appropriate tools and methodologies based on their specific analysis requirements and design goals.

\section{Memory Technologies in PIM Simulation}
\label{sec:mem_tech}

\begin{figure}[h]
\centering
\includegraphics[scale=0.4]{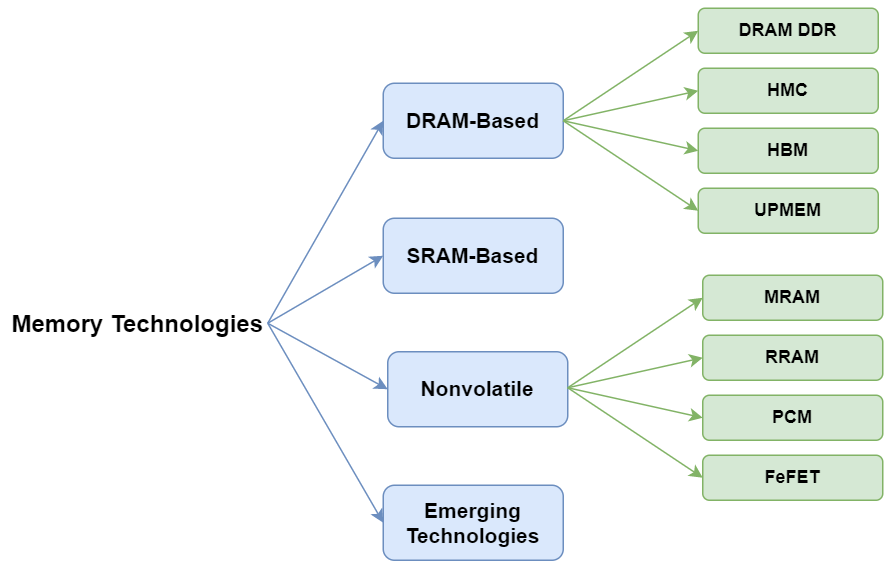}
\caption{Memory Technologies in PIM Simulation.}
\label{fig:mem_tech}
\end{figure}

PIM architectures leverage a diverse range of memory technologies, each offering distinct features that must be carefully considered during simulation. Unlike high-level architectural models or workload abstractions, the choice of memory technology fundamentally constrains the supported operations, achievable performance, and energy efficiency of a PIM system.
Consequently, memory should be treated as an independent and explicit simulation dimension rather than abstracted away within generic timing or workload assumptions \cite{khan2024landscapecomputenearmemorycomputeinmemoryresearch}. In this section, we highlight memory technology as a central axis of PIM simulation, emphasizing that properties of the underlying storage medium are as critical as the algorithms executed upon it. Figure~\ref{fig:mem_tech} and Table~\ref{tab:pim-memory} respectively illustrate the memory technologies and the representative simulators evaluated in this study.

\begin{table}[h]
\centering
\caption{Memory Technologies in PIM Simulation}
\label{tab:pim-memory}
\resizebox{\textwidth}{!}{
\begin{tabular}{|
    >{\centering\arraybackslash}m{5cm}|
    >{\arraybackslash}m{14cm}|
}
\hline
\textbf{Memory Technology} & \multicolumn{1}{c|}{\textbf{Representative Simulators}} \\
\hline
\textbf{DRAM (DDR)} &
NVMain, MPU-Sim, PIMSim, Ramulator-PIM, PiMulator, PIMeval, NDPmulator \\
\hline
\textbf{HBM} &
PIMSimulator, Ramulator-PIM, PiMulator, PIMeval \\
\hline
\textbf{HMC} &
SMCSim (SMC), MultiPIM, PIMSim, Ramulator-PIM, NDPmulator \\
\hline
\textbf{UPMEM} &
UPMEM SDK functional Sim, uPIMulator, UPMEM LLM Framework \\
\hline
\textbf{SRAM-based} &
MNSIM 2.0, NeuroSim, NDPmulator \\
\hline
\textbf{Non-Volatile Memories (NVM)} &
NVMain, NVSim, PIMSim, PIMSIM-NN, NeuroSim, MNSIM 2.0, 3D CiM LLM Sim, MemTorch \\
\hline
\textbf{Emerging Technologies} &
M$^2$NDP (CXL), AiM Simulator (CXL) \\
\hline
\end{tabular}
}
\end{table}

\subsection{DRAM-based PIM}
DRAM-based processing-in-memory architectures represent a prominent class of near-memory computing solutions, exploiting the widespread availability and maturity of DRAM technologies. These architectures integrate computational capabilities directly within or close to the DRAM arrays to reduce data movement and improve energy efficiency. DRAM-based PIM designs can be broadly categorized based on the underlying memory interface and organization. Traditional DDR-based PIM approaches extend conventional double data rate memory with lightweight processing units, while Hybrid Memory Cube (HMC) and High Bandwidth Memory (HBM) architectures adopt 3D-stacked DRAM with logic layers, enabling more sophisticated and tightly integrated PIM capabilities. Separately, UPMEM introduces a commercial PIM solution embedding general-purpose RISC-V cores within DRAM chips, offering a programmable and scalable platform for a range of memory-centric workloads. In the following subsections, we discuss the key characteristics, advantages, and challenges of DDR, HMC, HBM, and UPMEM-based PIM systems.

\subsubsection{DRAM DDR}
\begin{figure}[h]
\centering
\includegraphics[scale=0.4]{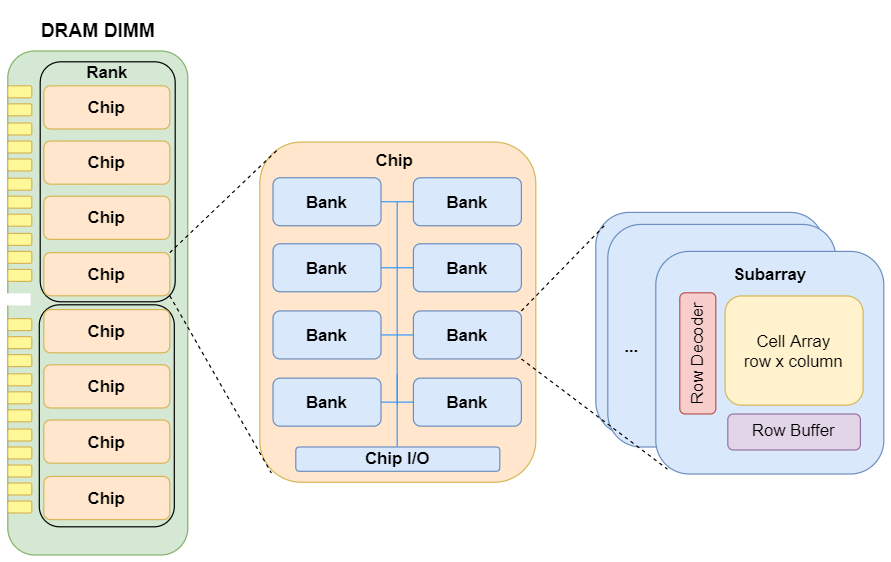}
\caption{Hierarchical organization of a DRAM module.}
\label{fig:DRAM_module}
\end{figure}

Modern DRAM is organized hierarchically, exemplified by DDR (Double Data Rate) DRAM, which is accessed through multiple memory channels operating concurrently. Each channel has its own address, data, and command buses. One or more dual in-line memory modules (DIMMs) can populate a channel; each DIMM contains one or more ranks, and a rank is a set of DRAM chips that each contributing a subset of bits for each data transfer.
Inside each chip, capacity and concurrency arise from multiple banks (typically 16 or more per chip); the same bank position across the chips in a rank forms a logical bank targeted by the memory controller \cite{10763591}.
Banks are further partitioned into subarrays, each with its own local row decoder and sense amplifiers to reduce electrical loading and improve timing. Hierarchical organization of a DRAM is shown in Figure~\ref{fig:DRAM_module}.

Despite this internal parallelism, the off-chip interface is comparatively narrow.
For DDR the global data lines (GDL) at the bank interface are typically 64 to 128 bits wide, and pinout and signaling constraints on the channel limit the effective parallelism that can be exploited \cite{10763591}.
Modeling studies commonly assume practical module configurations, for example, an x8 rank with eight chips, to accurately map the memory structure to bandwidth and capacity parameters \cite{khan2024landscapecomputenearmemorycomputeinmemoryresearch}.

For scale, many DRAM models used in PIM research assume around 16 banks per chip and dozens of subarrays per bank, with subarrays sized in the hundreds to a few thousand rows and several thousand columns. These assumptions provide simulators a realistic range of structural parallelism for mapping PIM kernels without committing to a specific vendor’s implementation \cite{10763591}.

Several well-known simulators model or emulate PIM behavior in DDR memories. \texttt{NVMain}, one of the earliest architectural-level simulators for emerging non-volatile and DRAM technologies, provides cycle-accurate modeling of timing, power, and hybrid DRAM–NVM configurations~\cite{Poremba2012NVMain}. Although originally designed for PCM and MRAM, its flexible controller and timing model make it a common baseline for DDR-based PIM studies. 

\texttt{PIMSim} builds directly on conventional DDR timing models (through DRAMSim2) and integrates programmable PIM cores within DRAM banks, allowing detailed evaluation of PIM execution, coherence, and data movement under standard DDR constraints~\cite{xu2018pimsim}. Similarly, \texttt{Ramulator-PIM}, an extension of the Ramulator memory simulator, adds in-memory compute units to DDR and other DRAM types~\cite{git2025ramulatorPIM}. It models the interactions between host processors and PIM cores through realistic DDR command traces, providing insight into bandwidth utilization and timing contention in 3D-stacked or multi-channel DDR environments.

Recent DDR-focused simulators have become increasingly detailed and heterogeneous. \texttt{PIMeval} introduces a unified performance and energy modeling framework for digital DRAM-PIM architectures, including subarray-level bit-serial, bit-parallel, and bank-level PIM designs within standard DDR4 configuration~\cite{10763591}. It provides a portable API and a benchmark suite that can evaluate different DDR-based architectures under uniform conditions.

\subsubsection{HMC}
\begin{figure}[h]
\centering
\includegraphics[scale=0.5]{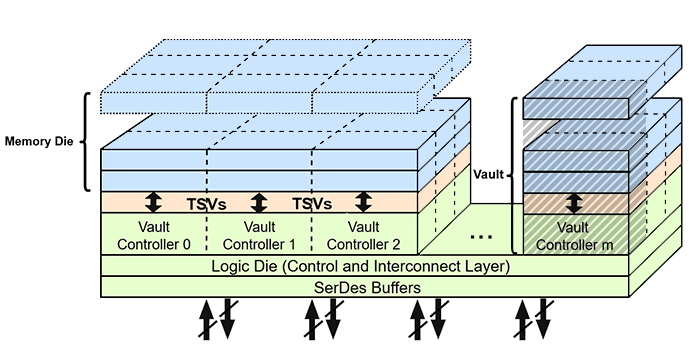}
\caption{HMC architecture \cite{asifuzzaman2023survey}.}
\label{fig:HMC_architecture}
\end{figure}

The Hybrid Memory Cube (HMC) is a 3D-stacked DRAM architecture comprising multiple DRAM layers stacked on top of a dedicated logic layer, connected vertically via through-silicon vias (TSVs) and micro-bumps, as illustrated in Figure~\ref{fig:HMC_architecture}.
The logic layer hosts the performance-critical control logic and partitions the stack into vertical “vaults,” each managed by an individual vault controller; a global controller coordinates vaults within the logic layer.
Since timing is managed internally within the device, the host interface is simplified compared to conventional DRAM.
HMC differs from High Bandwidth Memory (HBM) primarily in its host interface:
HMC exposes packet-based serial links optimized for CPU-friendly control and programming, while HBM uses wide parallel links over a silicon interposer, commonly paired with GPUs \cite{fujiki2021near}.

This organization makes HMC an ideal substrate for near-memory processing. 
The logic layer fabricated using a logic-optimized process, can host compute blocks without the performance penalties typical of DRAM-compatible logic.
Commonly, designs lightweight processing elements (PEs) are placed in the logic layer, one per vault, allowing each PE to access its local DRAM through short TSV paths.
A mesh Network-on-Chip (NoC) facilitates inter-vault communication when necessary.
This design achieves high internal bandwidth and reduces data-movement energy, though operations spanning distant vaults incur overhead due to remote traffic over the on-logic interconnect \cite{fujiki2021near}.

\texttt{MultiPIM}, a multi-stack PIM simulator models HMC-like stacked DRAM with vault controllers and internal networks. Each vault contains digital PIM cores executing memory-resident kernels. Computation is placed inside the vault logic layer, representing realistic near-bank processing within 3D-stacked DRAM cubes~\cite{Yu2021MultiPIM}. Through \texttt{PIMSim} configurable memory back-end, \texttt{PIMSim} can interface with HMCSim to emulate PIM in stacked DRAM~\cite{git2022PIMSim}. Compute logic remains in-vault, digital, while the HMC model supplies multi-vault bandwidth and fine timing. The simulator supports comparing PIM across DDR and HMC generations.

In \texttt{Ramulator-PIM}, By switching Ramulator’s memory configuration to HMC, the same near-memory PIM cores operate over 3D-stacked DRAM timing. Processing occurs in the logic die of the cube or at vault controllers. It remains a digital model but benefits from HMC’s internal parallelism~\cite{git2025ramulatorPIM}. \texttt{MPU-Sim} framework simulates massively parallel near-bank processors (MPUs) mapped onto HMC-like or stacked DRAM systems~\cite{Xie2022MPUSim}. Each vault hosts SIMT-style cores connected to the DRAM banks it manages. Compute thus resides near or within banks on the logic layer, exploiting stacked DRAM bandwidth.

The Smart Memory Cube (SMC) evolves the HMC concept by integrating programmable compute logic into the cube’s logic layer, transforming a passive memory controller into an active near-memory processing substrate \cite{Azarkhish2016SMCSim}.
Beyond vaults management, the logic die hosts lightweight processing elements that perform arithmetic and data-reduction operations directly within the memory stack, minimizing data movement to the host.
his enables fine-grained parallelism across vaults and exploits the cube’s high internal bandwidth for data-intensive workloads such as graph analytics and neural network inference. SMC thus serves as a practical intermediate step between conventional 3D-stacked DRAM and fully integrated PIM architectures, demonstrating how modest on-die compute capability can yield substantial energy efficiency and throughput gains \cite{Azarkhish2016SMCSim}.

\texttt{SMCSim} Designed for the Smart Memory Cube (SMC) platform, SMCSim reproduces full system behavior with compute elements embedded in the HMC logic die. Memory access obeys HMC vault timing, while compute instructions run in logic-layer digital cores. It bridges processor simulators and memory cubes to analyze near-memory workloads in stacked DRAM~\cite{Azarkhish2016SMCSim}.

\subsubsection{HBM}
High-Bandwidth Memory (HBM) is a 3D-stacked DRAM technology integrated side-by-side with a host ASIC (CPU, GPU, or accelerator) in the same package via a silicon interposer.
It stacks multiple DRAM layers atop a base logic die, connected through dense TSVs and micro-bumps and exposes an ultra-wide interface (typically around 1,024 I/Os per stack) that operates at moderate per-pin data rates (e.g., 3.2 Gb/s per pin for HBM2E).
This combination of vertical stacking and wide, short interconnect yields very high device bandwidth with superior I/O energy efficiency compared to traditional PCB-attached DRAM. Current stacks commonly comprise 4 to 8 layers, with 12-layer stacks planned, and system designs typically deploy 4 to 6 HBM devices around a large SoC \cite{micron2021hbm2e}.

Samsung’s \texttt{PIMSimulator} models the organization and operation of HBM-PIM systems. The simulator reflects the internal hierarchy of an HBM stack composed of several memory banks and vaults connected to a base logic die that hosts lightweight Processing Units (PUs). These PUs execute matrix and vector operations near the memory banks, following the real HBM2/HBM3 timing protocols. \texttt{PIMSimulator} allows researchers to configure and analyze architectural parameters such as the number of stacks, PUs per channel, command scheduling policies, and memory bandwidth allocation. By doing so, it captures both the compute-in-memory behavior and the host-PIM coordination that define HBM-PIM execution~\cite{git2025PIMSimulator}.

\subsubsection{UPMEM}
\begin{figure}[t]
\centering
\includegraphics[scale=0.08]{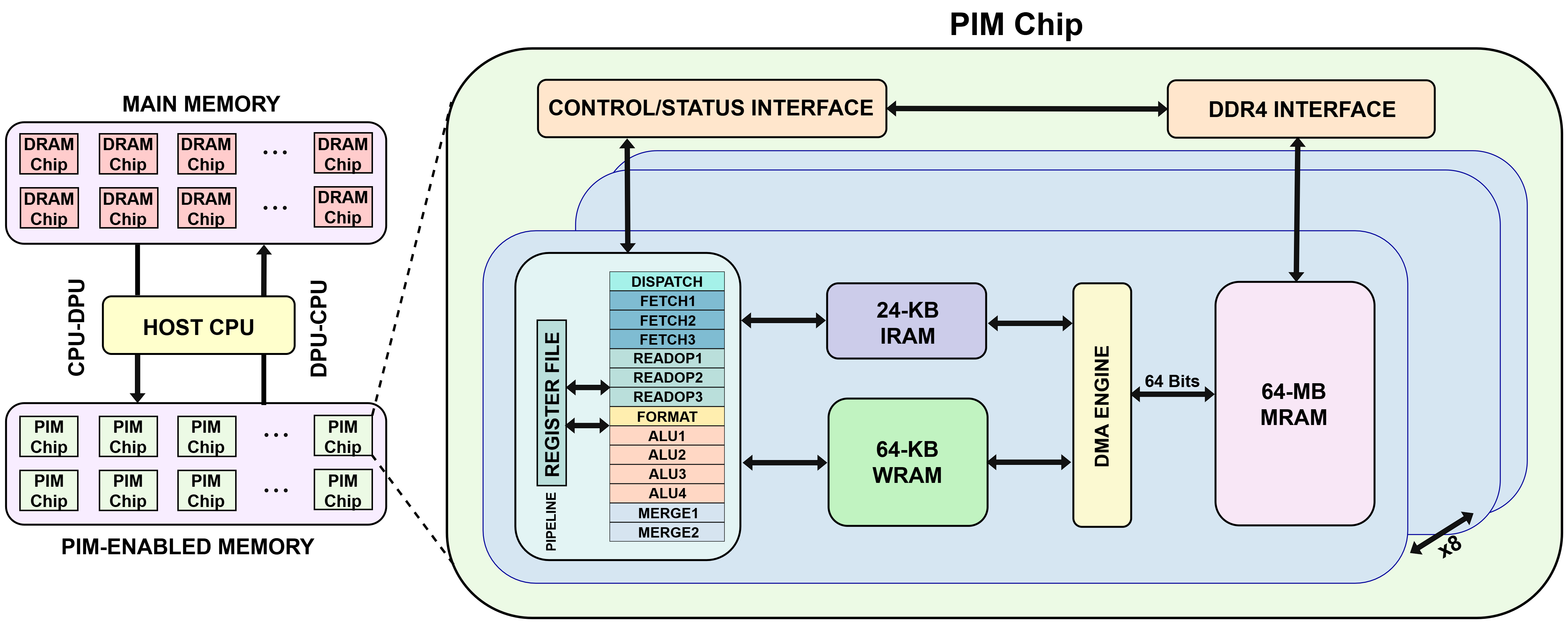}
\caption{Overview of the UPMEM architecture, illustrating the interaction between the host CPU, main memory, and PIM-enabled memory \cite{alonso2024bimsa}.}
\label{fig:UPMEM_architecture}
\end{figure}

UPMEM commercializes near-bank processing by integrating lightweight data processing units (DPUs) directly on the same die as commodity DDR4 DRAM (See Figure~\ref{fig:UPMEM_architecture}).
A typical PIM chip combines 4 Gb DDR4-2400 DRAM with eight DPUs running at approximately 500 MHz, and DIMMs are built from 16 such chips. UPMEM reports about 2 TB/s achieves an aggregate internal DRAM-DPU bandwidth of about 2 TB/s at that scale, while retaining a standard DDR4 interface and largely unmodified DRAM process technology \cite{8875680}.

Each DPU is a simple scalar, in-order, multithreaded core designed  to meet DRAM process constraints. To reach 500 MHz despite slow transistors, the microarchitecture uses a deep 14-stage pipeline and extensive fine-grained multithreading (24 hardware threads) to hide memory latency.

An autonomous DMA engine transfers code and data between DRAM (“MRAM”) and the on-chip SRAM. Programming follows a host-orchestrated model with a C/LLVM toolchain and runtime libraries; the CPU partitions data and dispatches kernels across thousands of DPUs \cite{8875680}.

\texttt{uPIMulator} is a cycle-accurate simulator of the commercial UPMEM DDR4-PIM architecture. Each DRAM chip integrates eight DPUs (14-stage in-order cores with IRAM/WRAM/MRAM)~\cite{Hyun2024uPIMulator}. Compute happens inside each DPU, physically located within the DRAM chip banks. Memory technology is standard DDR4 and processing is digital and fully programmable. It reproduces microarchitectural details—register files, pipelines, tasklets—validated against real hardware.

\texttt{UPMEM SDK functional simulator}, the functional simulator included in UPMEM’s official SDK emulates DPU behavior at the instruction level using real DDR4 timing profiles. It models in-DPU computation and standard DDR4 access delays but omits low-level electrical detail. The simulator is digital and serves as a fast functional tool for UPMEM programmers~\cite{Oliveira2024UPMEMSDK}.

The \texttt{UPMEM LLM Framework} extends the official UPMEM SDK and simulator to support large-scale evaluation of Large Language Models (LLMs) on UPMEM’s PIM architecture. Built on UPMEM’s instruction-level functional simulator, it enables efficient mapping, scheduling, and distributed execution of neural network layers across thousands of DRAM Processing Units (DPUs) in multiple DIMMs. The framework faithfully reproduces functional behavior, memory hierarchy, and bandwidth constraints of the real hardware, allowing accurate estimation of performance trends and scalability. By integrating directly with the UPMEM compiler and runtime environment, it provides a practical platform for exploring memory-centric AI workloads and optimizing model partitioning and data movement before deployment on physical PIM systems~\cite{git2024UPMEM_LLM_Framework}.

\subsection{SRAM-based PIM}
\begin{figure}[h]
\centering
\includegraphics[scale=0.35]{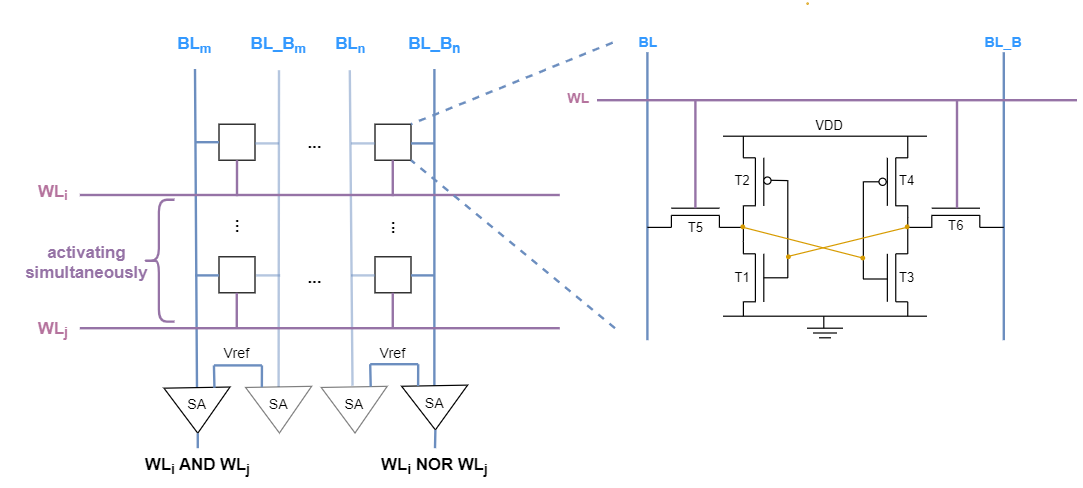}
\caption{Bitwise logical operations in SRAM and a calssic 6T SRAM cell~\cite{fujiki2021near}.}
\label{fig:SRAM_cell}
\end{figure}

Static RAM (SRAM) is the fast, on-chip workhorse of modern processor, storing bits using cross-coupled inverters (the classic 6T cell is shown in Figure~\ref{fig:SRAM_cell}). Unlike DRAM, SRAM does not require periodic refresh, enabling much lower access latency.
However, SRAM cells are larger and more complex, trading density and cost for speed. 
As a result, SRAM is typically used in small, high-performance structures such as register files and CPU caches near the top of the memory hierarchy \cite{fujiki2021near}.

SRAM-based PIM focuses on in-cache and SRAM-macro computing, where operations are placed inside or next to cache/memory subarrays to exploit bit-level parallelism and low-latency access. \texttt{NDPmulator} can attach Near-Data Accelerators (NDAccs) at any cache level (e.g., L1/L2) as well as DRAM, enabling realistic, full-system studies of in-cache/SRAM PIM alongside the OS and host CPU~\cite{Vieira2024NDPmulator}. Its scripts explicitly show NDAcc coupling to L2 and note that the same mechanism applies to L1 caches or DRAM, which is essential for modeling SRAM-resident PIM units within the cache hierarchy. \texttt{NDPmulator} also highlights prior in-cache processing work and contrasts it with its broader, multi-level support, reinforcing its suitability for SRAM-PIM evaluation across the hierarchy.

\texttt{MNSIM 2.0} provides a unified array model that captures both analog and digital PIM. critically, it includes SRAM-based digital PIM and validates against a fabricated dynamic-logic SRAM PIM macro. The paper reports low error versus silicon ($\approx$3.8\% for SRAM-digital PIM) and details readout via sense amplifiers and bitwise logic (e.g., AND) attached to memory cells, precisely the mechanisms used in many SRAM PIM macros. This lets authors sweep array granularity, activated rows/columns, and interface resolution while staying faithful to SRAM-specific peripheral circuits and timing~\cite{Zhu2023MNSIM2.0}.

Complementing this, \texttt{NeuroSim} models SRAM-centric compute-in-memory pipelines and peripherals, including SRAM cell scaling trends, SRAM buffers, and energy/leakage accounting, features needed to compare SRAM-based digital PIM against eNVM options~\cite{Lu2021NeuroSim}. Its benchmark setup explicitly includes the ADC requirements for 1-bit SRAM cells, enabling apples-to-apples assessments of accuracy/efficiency across device choices.

Together, NDPmulator (system/OS level), MNSIM 2.0 (behavioral/architectural with SRAM-digital validation), and NeuroSim (device/circuit-to-system with SRAM support) form a coherent toolchain for SRAM in your taxonomy, spanning from cache-level integration to per-array circuit effects. 

These same properties make SRAM an attractive substrate for processing-in-memory. Digital in-SRAM (“bitline”) schemes repurpose thousands of cache arrays as fine-grain, bit-serial arithmetic logic units (ALUs), effectively turning last-level caches into wide vector engines. This approach reduces overhead from cache controllers and on-chip networks, enabling high internal parallelism with one-cycle array accesses and minimizing data movement. Mixed-signal (analog) in-SRAM techniques go further by combining multi-row activation with lightweight DAC/ADC circuits, enabling vector-like operations directly on the shared bitlines. This can significantly boost energy efficiency and throughput when conversions costs are amortized \cite{fujiki2021near}.

\subsection{Non-Volatile Memories}
Non-volatile memories (NVMs) have emerged as promising technology for PIM architectures because they combine persistent data storage with the ability to perform computation directly within the memory array. Unlike DRAM, which requires refresh and destructive reads, many NVMs support non-destructive reads and can implement analog or digital logic in place.
This reduces costly data movement and enables massive internal parallelism, offering energy-efficient computing beyond the traditional von Neumann model.

Some of prominent NVM technologies include magnetoresistive RAM (MRAM), resistive RAM (RRAM), phase-change memory (PCM), and ferroelectric FETs (FeFETs), each relying on distinct physical switching mechanisms. While these technologies offer advantages such as non-volatility, scalability, and compatibility with CMOS processes, they also face challenges such as endurance limits, variability, and write latency/energy trade-offs that must be addressed for practical large-scale systems \cite{jahannia2024multi, cheshmikhani2019robin}.

\subsubsection{MRAM}
Magnetic RAM (MRAM) cells are built around a magnetic tunnel junctions (MTJs), which consist of two ferromagnetic layers separated by a thin tunnel barrier.
MTJ is a metal–insulator–metal (MIM) stack that consists of two ferromagnets separated by a thin tunnel barrier.
One layer (``reference'') has a fixed magnetization, while the other (``free or storage'') layer is switchable \cite{7555318}. The MTJ resistance depends on the relative relative orientation of these layers: parallel alignment yields a low-resistance state; anti parallel alignment yields a high-resistance state.
Creating this relative orientation of the magnetizations of layers is known as tunneling magnetoresistance (TMR) effect.
In practice, the free layer, tunnel barrier (often ~1 nm MgO), and reference layer are co-designed to provide two well-separated, stable resistance states representing binary data \cite{7555318, farbeh2016floating}.

MRAM arrays typically use a compact 1T–1MTJ cell, where a select transistor gates read and write currents. MRAM combines non-volatility, high-speed read/write, and CMOS compatibility. However, designers must manage bit-to-bit resistance variation and scaling challenges that tighten read margins as devices shrink \cite{maurer2025survey}.

\subsubsection{RRAM}
\begin{figure}[h]
\centering
\includegraphics[scale=0.4]{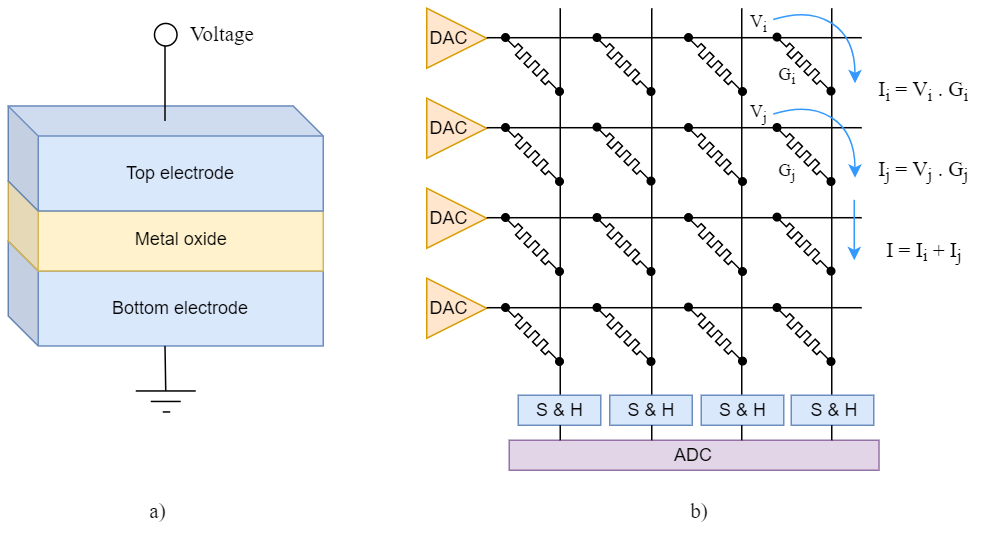}
\caption{Nonvolatile memories: a) an RRAM cell, b) in place vector-matrix multiplication \cite{dorostkar2024empirical}.}
\label{fig:nvm}
\end{figure}

Resistor RAM (RRAM) stores information by modulating the resistance of a simple MIM stack, typically a thin metal-oxide layer sandwiched between electrode. An applied electric field creates and ruptures conductive filaments of oxygen vacancies within the oxide switching the device between a high-resistance state (HRS) and a low-resistance state (LRS) (SET and RESET operations, respectively)~\cite{7950949}.
Filament dynamics are stochastic, causing SET/RESET thresholds to vary across write cycles and devices. Figure~\ref{fig:nvm} a) depicts an RRAM cell, while Figure~\ref{fig:nvm} b) demonstrates its use for in place vector–matrix multiplication.

RRAM is attractive for tightly integrated PIM because it can be fabricated in back-end-of-line (BEOL) interconnect layers with CMOS-compatible materials and stacked monolithically in 3D \cite{7950949, ghasemi2022grapha}.
While RRAM offers strong scaling potential and integration benefits, variability in switching thresholds and leakage in low-resistance states remain challenges, requiring adaptive write/read strategies and static power management \cite{maurer2025survey}.

\subsubsection{PCM}
Phase-change memory (PCM) stores bits by switching a chalcogenide material between a high-resistance amorphous phase and a low-resistance crystalline phase, enabling significant read-current contrast and even multi-level storage \cite{Burr_2010}.
Writing involves two thermal modes: SET recrystallizes the material by heating above the crystallization temperature, while RESET melts and rapidly quenches it to the amorphous state \cite{Burr_2010}.

SET operations are typically slower nucleation and crystal growth, limiting write throughput.
At the same time, PCM’s high operating temperatures and large RESET current complicate fabrication and scaling, as access devices must supply sufficient current interfaces must withstand thermal stress.
Despite fast switching and strong read contrast, PCM designers must balance RESET power/speed with SET-limited write performance under tight thermal and process constraints \cite{maurer2025survey}.

\subsubsection{FeFET}
Ferroelectric FETs (FeFET) integrate a ferroelectric oxide layer into the MOSFET gate stack.
Non-volatility arises from polarization hysteresis in the ferroelectric layer, with two stable threshold-voltage states encoding binary data \cite{10.1145/3218603.3218640}.
FeFETs separate read and write paths, with write operations flipping ferroelectric polarization via gate-to-source voltage without current flowing through the drain–source channel.
Voltage-driven writes result in lower operating energy compared to current-driven two-terminal NVMs such as RRAM, and FeFETs offer strong retention. However, endurance remains a significant concern for large-scale deployment \cite{maurer2025survey}.\\

\texttt{NVSim} provides parametric modeling for emerging non-volatile memories such as PCM, RRAM, MRAM, and FeFET. While it doesn’t execute compute, it quantifies latency, energy, and area of arrays that can later serve as analog or digital PIM blocks~\cite{Dong2012NVSim}. \texttt{NVMain} Complements NVSim with cycle-accurate timing for PCM, RRAM, and hybrid DRAM–NVM. It remains primarily a memory behavior simulator, but researchers have used it to attach PIM controllers to banks for near-array digital compute~\cite{Poremba2012NVMain}.

\texttt{NeuroSim} models analog in-memory computation using MRAM, RRAM, PCM, and FeFET crossbars. Computation is in-array analog multiply-accumulate (MVM) with peripheral ADC/DAC circuits. It spans device-to-architecture levels, capturing non-idealities like variation and quantization~\cite{Lu2021NeuroSim}. \texttt{MNSIM 2.0} combining analog RRAM Computing In Memory and digital SRAM PIM in one framework. Computation can occur analog inside crossbar arrays or near-array digital units depending on configuration~\cite{Zhu2023MNSIM2.0}.

\texttt{MemTorch} integrates PyTorch with models of RRAM memristive crossbars. It simulates in-array analog computation including stochastic and conductance-drift effects, allowing neural network training and inference analysis under realistic NVM behavior~\cite{Lammie2022MemTorch}. \texttt{PIMSIM-NN}, An ISA-based, cycle-accurate simulator for RRAM memristor-based PIM accelerators. It uses SystemC to orchestrate matrix (crossbar), vector, and scalar units. Computation occurs in-array analog MVM managed by digital control units. It supports ONNX models and models full chip timing~\cite{10546788}. IBM \texttt{3D-CiM LLM Inference Simulator} for 3D analog in-memory computing, aimed at LLM inference. It abstracts multi-tier PCM/RRAM crossbars stacked vertically. Users supply latency/energy parameters for each analog operation. Processing is in-array analog, across multiple 3D layers, representing IBM’s PCM-based AIMC research line~\cite{git2025CiM3DLLMInference}.

\subsection{Emerging Technologies}
\begin{figure}[h]
\centering
\includegraphics[scale=0.09]{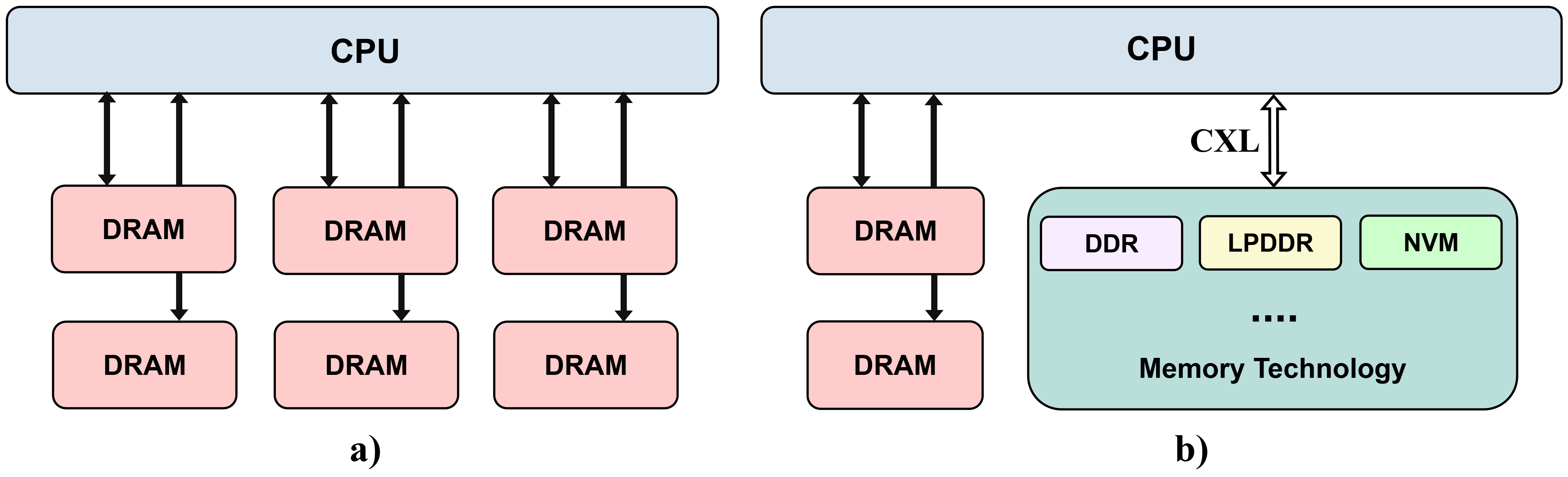}
\caption{CXL attached memory \cite{maruf2023tpp}: a) without CXL, b) with CXL.} \label{fig:CXL}
\end{figure}

Compute Express Link (CXL) is not a memory technology, but a cache-coherent, high-bandwidth interconnect that enables memory devices and accelerators attach to the CPUs as coherent peers, as shown in Figure~\ref{fig:CXL}.
For PIM, CXL facilitates packaging intelligent memory modules directly into the system fabric as first-class devices, enabling external memory-side processors that share coherent data structures with the CPU. This approach simplifies software integration compared to traditional DMA-based offload but introduces fabric latency ($\sim$100--200 ns round-trip) and bandwidth constraints compared to native DDR access \cite{10764494}.

A forward-looking PIM simulator should model CXL link latency and bandwidth, protocol overheads including coherence traffic (such as cache invalidations/updates caused by PIM operations), and fabric topologies that enable distributed PIM modules cooperating over a switch. Programming models can treat such PIM modules as coherent peers or memory-mapped devices, reducing software complexity. Thermal/reliability concerns shift toward the interconnect components rather than memory cells, enabling disaggregated, pooled memory with integrated compute and expanding architectural possibilities for PIM.

\texttt{M$^2$NDP} models Near-Data Processing inside CXL.memory expanders backed by DRAM (typically LPDDR5). Compute engines reside within the CXL device controller near the DRAM channels, using coherent CXL.mem protocols. Processing is digital and near-memory rather than in-cell~\cite{10764494}.

\texttt{AIM Simulator} models a GDDR6-PIM system connected via CXL. Each DRAM bank embeds digital MAC arrays, while near-memory units in the CXL controller handle higher-level operations. Computation is split between in-bank PIM and near-memory digital cores. Memory technology is GDDR6 DRAM with modest logic integration~\cite{Gu2025AiMsimulator}.


\section{Application Domains of PIM Simulators}
\label{sec:sim_domains}

\begin{figure}[h]
\centering
\includegraphics[scale=0.4]{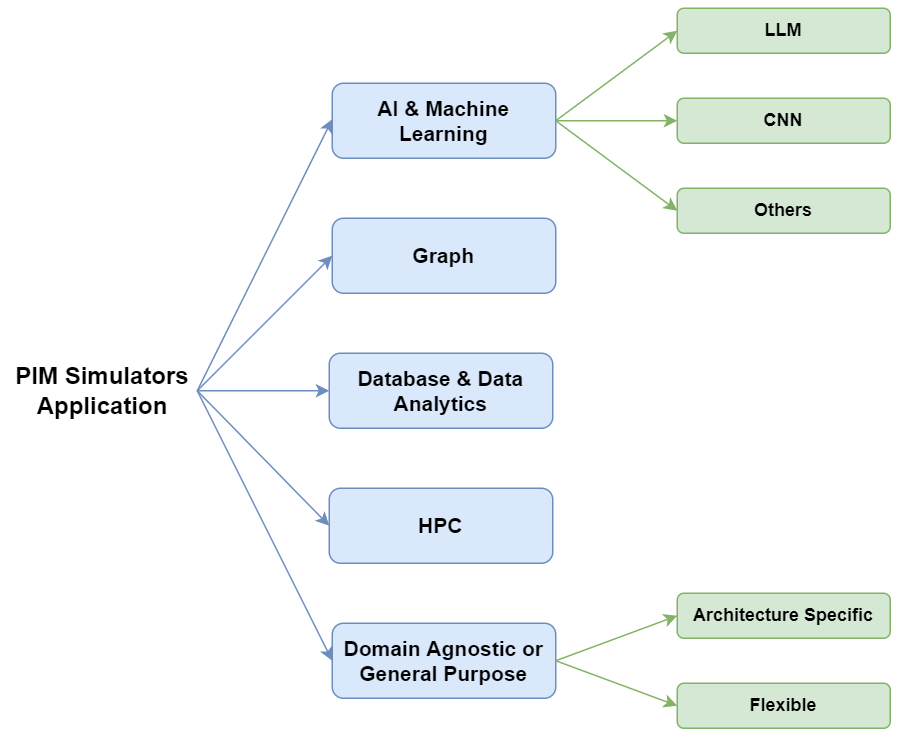}
\caption{PIM Simulators Applications Classification}
\label{fig:pim_domains}
\end{figure}

PIM simulators have been evaluated across various application domains. Their classification based on application and usage largely hinges on the types of benchmarks employed in their validation studies or their overall use cases. Some simulators are highly domain-specific, concentrating workloads such as deep learning or graph analytics, while others are designed as general-purpose frameworks to accommodate a wide array of applications. In this section, we categorize PIM simulators into five primary groups: 1) AI and Machine Learning, 2) Graph Analytics, 3) Database and Data Analytics, 4) High-Performance Computing (HPC), and 5) Domain Agnostic or General Purpose. The visual overview of this classification is presented in Figure~\ref{fig:pim_domains}, along with Table~\ref{tab:pim-domains}, which illustrate representative PIM simulators.

\begin{table}[t]
\centering
\caption{Application Domains of PIM Simulators}
\label{tab:pim-domains}
\resizebox{\textwidth}{!}{
\begin{tabular}{|
    >{\centering\arraybackslash}m{6cm}|
    >{\arraybackslash}m{13cm}|
}
\hline
\textbf{Application Domain} & \multicolumn{1}{c|}{\textbf{Representative Simulators}} \\
\hline
\textbf{AI / Machine Learning} &
PIMSim, PIMeval, MPU-Sim, PIMSim-NN, NeuroSim V2.1, MNSIM 2.0, uPIMulator, MemTorch, NDPmulator, M$^2$NDP, 3D CiM LLM Sim, PIMSimulator, AiM Simulator, UPMEM SDK functional Sim, UPMEM LLM Framework \\
\hline
\textbf{Graph Analytics} &
SMCSim, PIMeval, PIMSim, Ramulator-PIM, MPU-Sim, MultiPIM, uPIMulator, M$^2$NDP \\
\hline
\textbf{Database / Data Analytics} &
PIMeval, PIMSim, UPMEM SDK functional Sim, MPU-Sim, uPIMulator, NDPmulator, M$^2$NDP \\
\hline
\textbf{High-Performance Computing} &
NVMain, SMCSim, PIMSim, PIMeval, Ramulator-PIM, MPU-Sim, PIMSimulator, uPIMulator, NDPmulator, M$^2$NDP, MultiPIM \\
\hline
\textbf{General Purpose} &
uPIMulator, SMCSim, Ramulator-PIM, MPU-Sim, PIMeval, PIMSim, PiMulator, NDPmulator, M$^2$NDP, PIMSimulator, MultiPIM, NVMain, UPMEM SDK functional Sim, UPMEM LLM Framework \\
\hline
\end{tabular}
}
\end{table}

\subsection{AI and Machine learning}
AI workloads, especially deep learning, are often memory-bound. Both training and inference involve massive matrix–vector multiplications (MVMs), convolutions, and tensor transformations. In GPUs and CPUs, moving data from DRAM to compute units dominates energy and latency: a single 64-bit DRAM access can consume roughly 1000 times more energy than a floating-point multiply~\cite{10.1145/3695794.3695797}.
AI workloads impose distinct computational and memory access patterns during training and inference, each presenting unique challenges to system design. Moreover, the presence of both dense and sparse operations across various tensor formats necessitates flexible and efficient dataflow mechanisms~\cite{sze2017efficient}. The critical characteristics of these workloads can be summarized as follows:
\begin{itemize}
    \item \textit{Training:} Requires frequent weight updates, backpropagation, and gradient storage, which generate enormous read–write traffic between compute and memory \cite{jouppi2023tpu}.
    \item \textit{Inference:} Models like CNNs, transformers, and large language models (LLMs) need large parameter fetches and key–value (KV) caches for attention, stressing bandwidth and memory capacity \cite{yu2022orca}.
    \item \textit{Sparsity \& Tensor Diversity:} Both dense (e.g., matrix multiplications in CNNs) and sparse (e.g., attention mechanism, recommendation systems) operations occur, demanding flexible dataflows \cite{choquette2021nvidia}.
\end{itemize}

Most simulators in this category focus on the accelerating General Matrix-Matrix Multiplication (e.g, \texttt{PIMSIM-NN}~\cite{git2024PimSimNN}, \texttt{AIHWKit}~\cite{git2025AIHWKIT}) and General Matrix-Vector Multiplication (e.g, \texttt{PIMeval/PIMbench}~\cite{git2025PIMevalPIMbench}).
As a result, PIM helps because:
a) avoids costly data transfers caused by heavy computations such as multiply-accumulate (MAC), pooling, and attention kernels; b) scales bandwidth with memory capacity; c) use emerging memories (ReRAM, PCM, FeFET) that natively support analog MVMs \cite{spiga2020memristive}, well aligned with AI’s linear algebra backbone.


\subsection{Graph}
Graph processing algorithms involve irregular memory access, pointer chasing, and low arithmetic intensity. Operations such as neighbor traversal, frontier expansion, and vertex updates require fetching scattered data across large memory spaces \cite{ahn2015scalable}. Unlike AI workloads that are compute-intensive, graph workloads are often limited by random memory latency. 
Critical operations in Graphs include:
\begin{itemize}
    \item \textit{Breadth-First Search (BFS):} repeated queue/frontier expansion, heavy on atomic updates \cite{zhang2018graphp}.
    \item \textit{PageRank:} iterative floating-point updates to vertex ranks.
    \item \textit{Shortest Path (Dijkstra/Bellman-Ford):} frequent check-and-update operations on edges.
    \item \textit{Graph Traversals (DFS, SSSP, Connected Components):} pointer chasing across adjacency lists.
\end{itemize}

These operations cause bandwidth bottlenecks and high synchronization costs. CPUs and GPUs often stall on random DRAM accesses, while PIM architectures reduce traffic by performing updates (atomics, comparisons, rank increments) directly inside memory \cite{7529923}. Currently, many PIM architectures are proposed to ease and speedup graph processing, such as Tesseract \cite{ahn2015scalable}, GraphP \cite{zhang2018graphp}, and GraphH \cite{dai2018graphh}. Also, simulator like \texttt{HMC-Sim} \cite{7529923}, and \texttt{PIMeval} \cite{git2025PIMevalPIMbench} support graph processing workloads. 


\subsection{Database and Data Analytics}
Database management systems (DBMS) and analytics platforms are dominated by data movement rather than arithmetic computation. Typical queries such as joins, aggregations, scans, filters, require streaming or shuffling gigabytes of tuples across memory hierarchies. For example, TPC-H benchmarks show that selection, projection, join, and aggregation account for nearly 90\% of total execution time and memory traffic\cite{10.14778/3368289.3368298}.

On conventional CPUs and GPUs, data must travel from DRAM through caches into execution units, even if only simple comparisons or additions are performed. PIM mitigates this by executing common database primitives such as comparison, filtering, aggregation, and sorting, directly inside or near DRAM banks. This reduces latency, energy consumption and intermediate traffic. 
Critical Database Operations for PIM are as follows:
\begin{itemize}
    \item \textit{Selection/Filtering:} ideal for SIMD-style execution inside DRAM vaults.
    \item \textit{Projection:} can be combined with filtering to reduce result size.
    \item \textit{Aggregation and Grouping:} bandwidth-bound; atomic add/multiply inside memory reduces traffic.
    \item \textit{Join:} extremely data-intensive; benefits from PIM’s parallel row scans and bulk bitwise operations.
    \item \textit{Sorting:} bandwidth- and compute-intensive; PIM can accelerate sorting primitives. 
\end{itemize}

To address the need for PIM databases, architectures such as \texttt{PIM-enabled instruction sets} \cite{ahn2015pim} are proposed to show that common database operations like selection and aggregation can be executed within a 3D-stacked memory logic layer, avoiding the round trip to the CPU for simple comparisons and arithmetic. This study supports in-memory hash join, histogram and radix partitioning. In addition, \texttt{PIMeval/PIMbench} \cite{git2025PIMevalPIMbench} specifically supports database and data analysis benchmarks. 


\subsection{High Performance Computing (HPC)}
High-performance computing workloads are both memory- and compute-intensive requiring sustained high bandwidth, low latency access, and scalability across thousands of cores.
Critical HPC operations include:
\begin{itemize}
    \item Stencil Updates: iterative computations on structured grids \cite{denzler2023casper}.
    \item Sparse Matrix-Vector Multiplication (SpMV) \cite{zhang2014top}.
    \item Fast Fourier Transform(FFT) and spectral transforms \cite{ibrahim2024pimacolaba}.
    \item Partial Differential Equation(PDE) solvers (wave propagation, elasticity).
    \item Linear Algebra Kernels (BLAS, LU/Cholesky factorization).
\end{itemize}

Traditional HPC nodes suffer from the memory wall, where interconnect and DRAM bandwidth cannot keep pace with core scaling. PIM simulators for HPC explore how HBM or HMC with in-memory compute can efficiently reduce traffic for these kernels.

Some of the research proposals in HPC include \texttt{Casper} \cite{denzler2023casper}, which accelerates stencil computations with specialized units placed near the CPU's last-level cache to maximize local bandwidth. Other work like \texttt{TOP-PIM} \cite{zhang2014top} embeds programmable, throughput-oriented GPU cores within 3D-stacked memory to serve as PIM accelerators for kernels including SpMV. For bandwidth-bound tasks like FFTs, collaborative approaches such as \texttt{Pimacolaba} \cite{ibrahim2024pimacolaba} are proposed, splitting the computation between the host GPU and PIM hardware to balance high bandwidth with compute power, thereby improving performance and reducing data movement.
Simulators such as \texttt{PIMSimulator} \cite{git2025PIMSimulator}, \texttt{HMC-Sim} \cite{7529923}, and \texttt{PIMeval/PIMbench} \cite{git2025PIMevalPIMbench} support HPC-related workloads. 


\subsection{Domain Agnostic or General Purpose}
These simulators evaluate a broad spectrum of workloads (AI, HPC, graph, database, analytics) rather than focusing on a single domain. Some are flexible frameworks (e.g., \texttt{UniNDP}~\cite{xie2025unindp}); others are architecture-specific (e.g., \texttt{MIMDRAM}~\cite{oliveira2024mimdram}, \texttt{AttAcc}~\cite{park2024attacc}, \texttt{NeuPIMs}~\cite{heo2024neupims}). They enable fair comparisons across PIM designs and system-level integration and exploration. 
Some are based on real architectures (e.g., \texttt{UPMEM SDK}~\cite{Oliveira2024UPMEMSDK}, \texttt{uPIMulator~\cite{git2025uPIMulator}}), while others are abstract research tools providing configurable backends (e.g., \texttt{PIMSim}~\cite{xu2018pimsim}, \texttt{MultiPIM}~\cite{Yu2021MultiPIM}).



\section{Contexts of PIM Simulator Deployment}
\label{sec:sim_deployment}

PIM simulators today play diverse roles across academic research, industrial design, and education. While most tools originate within one of these domains, their applications often extend well beyond their initial scope.
For example, academic simulators such as \texttt{NeuroSim} and \texttt{NVSim} \cite{Lu2021NeuroSim, Dong2012NVSim} have evolved into widely adopted industrial frameworks, while industrial simulators like \texttt{UPMEM’s SDK} are employed in university research and training purposes.
Therefore, PIM simulators should not be regarded as exclusive to a single context; rather, they form a flexible ecosystem that adapts to diverse needs including architectural exploration, design validation, and education purposes.   
Broadly, three deployment contexts can be identified:
\begin{itemize}
    \item Academic research focused on exploring new architectures and evaluating untested ideas.
    \item Industrial development aimed at product validation, pre-silicon optimization, and reliability assessment.
    \item Education and training dedicated to teaching, demonstration, and skills development.
\end{itemize}

Each context imposes specific requirements on simulation frameworks, including accuracy, scalability, usability, and integration with existing toolchains.
The following sections discuss these contexts in detail and highlight representative simulators in each category.

\subsection{Academic Research Applications}
In academic research, PIM simulators serve as essential platforms for investigating architectures that bring computation closer to memory. They bridge the gap between theoretical design and hardware prototyping by providing controlled, low-cost, and reproducible environment for design-space exploration, performance analysis, and hardware–software co-design.

Modern academic simulators, such as \texttt{NDPmulator} and \texttt{MultiPIM}, offer system-level emulation of CPUs, memory hierarchies, and near-data accelerators under both user-space and kernel-space conditions \cite{Vieira2024NDPmulator, Yu2021MultiPIM}.

Cycle-accurate tools like \texttt{PIMSIM-NN} and \texttt{MNSIM 2.0} enable researchers to evaluate neural-network workloads with various quantization schemes, mapping strategies, and fault-tolerance mechanisms \cite{10546788, Zhu2023MNSIM2.0}.

A key recent trend is cross-layer simulation, where architectural frameworks integrate detailed circuit- or device- models.
For example, \texttt{NeuroSim} and \texttt{NVSim} provide validated circuit- and device-level models that are often linked with architecture-level tools like \texttt{PIMulator-NN} or \texttt{MNSIM 2.0}. This integration enables end-to-end analysis of non-idealities, power consumption, and accuracy degradation in both digital and analog PIM designs \cite{Lu2021NeuroSim, Dong2012NVSim}.

At higher abstraction levels, simulators such as \texttt{PIMSim} and \texttt{Ramulator-PIM} support flexible configuration of DRAM timing parameters, interconnect typologies, and offloading policies. These frameworks are widely used to evaluate PIM architectures \cite{xu2018pimsim, git2025ramulatorPIM} and also CXL-based NDP architectures, as exemplified by \texttt{M$^2$NDP} \cite{10764494}. 

Thus, academic simulators provide a foundation for innovation by lowering experimental barriers. They enable research groups to evaluate emerging paradigms, such as 3D-stacked DRAM PIM, ReRAM-based analog computing, or hybrid CXL–PIM architectures, well before hardware fabrication, accelerating discovery in the field.

\subsection{Industrial Applications}
In industry, PIM simulators shift from exploration to verification, optimization, and integration within product design workflow. These tools emphasize cycle accuracy, determinism, and compatibility with commercial EDA or AI frameworks, serving as essential pre-silicon validation platforms.

A prominent examples is the \texttt{UPMEM SDK Functional Simulator}, part of the official development kit for UPMEM’s commercial DRAM-PIM chips \cite{Oliveira2024UPMEMSDK}. It accurately models real DPU (Data Processing Unit) behavior, enabling developers to compile, debug, and profile applications using the same APIs as on hardware. This simulator underpins both industrial and academic collaborations focused on programming models and workload characterization.

\texttt{UPMEM’s PIM-AI} framework extends this concept to Large Language Model (LLM) inference \cite{ortega2024PIM-AI} integrating PyTorch-level workloads with realistic models of energy consumption, throughput, and total cost of ownership (TCO). This allows accurate performance estimation under cloud or mobile deployment scenarios.
Similarly, \texttt{uPIMulator} offers cycle-level simulation of commercial UPMEM-PIM systems. Calibrated against hardware measurements, it has become a reference model for evaluating commercial DRAM-PIM technologies within the research community \cite{Hyun2024uPIMulator}.

Samsung’s \texttt{PIMSimulator}, developed by the Samsung Advanced Institute of Technology (SAIT), provides a cycle-accurate model of HBM and logic-layer interactions \cite{git2025PIMSimulator}, helping engineers in analyzing PIM instruction sets and memory behavior before silicon implementation.
Industrial R\&D centers have introduced simulators for specialized use cases as well. The \texttt{IBM 3D-CiM LLM Inference Simulator} models hybrid analog–digital compute-in-memory architectures with detailed 3D stacking and thermal considerations, integrated into \texttt{IBM’s AIHWKit} and chip design workflows \cite{Buchel2025EfficientScaling, git2025AIHWKIT}.

Several academic–industrial hybrid frameworks further blur the boundaries between research and product development. 
For instance, \texttt{PiMulator} is an FPGA-based emulation platform achieving up to 28× faster evaluation than software simulators, providing near-real-time prototyping \cite{Mosanu2022PiMulator}. Similarly, \texttt{M$^2$NDP} (POSTECH \& SK hynix) enables cycle-accurate modeling of CXL-based memory expanders with embedded computation \cite{10764494}.

At the device and circuit level, mature tools such as \texttt{NeuroSim} and \texttt{NVSim} remain cornerstones for both academia and industry, serving semiconductor vendors like Intel, SK hynix, and TSMC for technology evaluation \cite{Lu2021NeuroSim, Dong2012NVSim}.
Overall, industrial PIM simulators focus on hardware prototyping, product validation, and design optimization. They provide deterministic, performance-calibrated models that shorten design cycles, reduce development cost, and mitigate fabrication risks.

\subsection{Educational and Training Applications}
Beyond research and product design, PIM simulators have emerged as valuable tools for education and professional training. They enable learners to observe the interaction of computation and data within memory hierarchies, connecting theoretical concepts to measurable system behavior.

Simplified frameworks such as \texttt{DRAMSim3}, \texttt{Ramulator-PIM}, and \texttt{PIMSim} are used in undergraduate courses to visualize memory operations like row activation, bank-level parallelism, and near-data execution \cite{Li2020DRAMsim3, git2025ramulatorPIM, xu2018pimsim}. By conducting simple experiments, students can directly observe how PIM reduces data movement and improves energy efficiency.

At the graduate and research level, tools like \texttt{MemTorch}, \texttt{NeuroSim}, and \texttt{PIMSIM-NN} facilitate hands-on exploration of neuromorphic and compute-in-memory accelerators. Their Python and PyTorch integration make them ideal for understanding co-design principles, circuit variations, and algorithmic resilience \cite{Lammie2022MemTorch, Lu2021NeuroSim, 10546788}.

Industrial frameworks, including \texttt{UPMEM’s SDK} and \texttt{IBM’s 3D-CiM simulator}, are also valuable for corporate training, helping engineers gain familiarity with data-centric computing and heterogeneous memory systems.

These environments provide realistic experimentation without requiring physical access to hardware, making them ideal for skill development and technology adoption. Educational deployment of PIM simulators thus bridges theory and practice, preparing a new generation of architects and engineers capable of integration software insight with hardware innovation.


\section{Evaluation Metrics for PIM Simulation}
\label{sec:sim_metrics}
The rigorous evaluation of modern memory architectures, whether conventional DRAM, emerging NVM, or tightly coupled Processing-in-Memory accelerators requires a comprehensive set of metrics. These metrics range from foundational processor performance indicators such as Clock Per Instruction (CPI) to specialized characteristics related to power, area, and reliability specific to the underlying memory technology. Simulator frameworks integrate these metrics across multiple abstraction levels, from device physics to the application layer, to enable efficient design space exploration.

\subsection{Performance Metrics}
Performance metrics remain the cornerstone of architectural evaluation. They quantify how quickly and effectively a system executes workloads and processes data.
These metrics encompass aspects such as latency, throughput, and bandwidth, each essential for characterizing overall system behvior and identifying the bottlenecks.

\subsubsection{Core Latency and Execution Efficiency}
This category of metrics measures the fundamental speed of the processing elements, ranging from host CPU to PIM cores.It also accounts for memory access latency and communication latency across interconnects. 

\vspace{5pt}
\noindent
\emph{Execution Time and IPC/CPI:}
The most fundamental metric used to assess performance is \textit{execution~time} (T), which is extracted as follows:\\
\[T = N \times CPI \times \frac{1}{f}\]\\
where \textit{N} is the total number of executed instructions, \textit{CPI} represents cycles per instruction, and \textit{f} is the operating clock frequency \cite{10.5555/1855084}.

Closely related metrics such as Instructions per Cycle (IPC) and its reciprocal, CPI, highlight utilization efficiency of the execution pipeline. CPI breakdowns are used to isolate delays due to stalls, cache misses, or pipeline hazards \cite{10.5555/1855084}.

In PIM architectures, total latency is often decomposed into \textit{data movement latency}: the time spent transferring data between the CPU and memory,  and \textit{PIM kernel execution latency}: the time spent executing computational operations directly within memory subsystems \cite{git2025PIMevalPIMbench}.

\vspace{5pt}
\noindent
\emph{Speedup:}
Speedup is a universal metric used to express the performance improvement achieved by a baseline (e.g., CPU or GPU).

The theoretical limits of speedup are dictated by Amdahl’s Law, which states that the improvement is bounded by the fraction of the program that can be parallelized. Therefore, only the parallelizable components of a workload, such as PIM-accelerated computation or optimized DRAM access, contribute to meaningful speedup \cite{10.5555/1855084}.

\vspace{5pt}
\noindent
\emph{Latency Stacks and Timing Breakdown:}
Simulators often employ latency stack diagrams to visualize the cumulative delays in workload execution. These stacks typically include, queuing, DRAM access, and compute latency.
Such visualizations help in identifying dominant bottlenecks in the system and are critical for architectural tuning and optimization.

\subsubsection{Memory Bandwidth and Throughput Analysis}
These metrics are tailored to evaluate the performance of the memory subsystem, particularly in memory-bound applications where the memory wall becomes the limiting factor.

\vspace{5pt}
\noindent
\emph{Bandwidth Utilization:}
Bandwidth utilization measures the ratio of achieved memory bandwidth to the theoretical peak bandwidth.
Low utilization in a memory-intensive workloads highlights bottlenecks in the data path, interconnect, or memory controller scheduling.

\vspace{5pt}
\noindent
\emph{Workload Parallelism and Scaling (STP/ANTT):}
In systems supporting multiprogramming or multithreading with shared DRAM access, metrics such as System Throughput (STP) and Average Normalized Turnaround Time (ANTT) are vital. STP quantifies the aggregate throughput across concurrent tasks, while ANTT reflects fairness and latency per workload.

These metrics are critical for characterizing DRAM access conflicts, the impact of non-determinism, revealing the trade-offs between maximizing overall bandwidth usage and ensuring fair resource allocation across concurrent tasks \cite{10.5555/1855084}.

\subsection{Power and Energy Metrics}
As data movement becomes the dominant contributor to system-wide power consumption, energy metrics have grown in importance, often becoming the ultimate constraint for designing scalable and sustainable architectures.
These metrics span from low-level circuit energy use to application-level energy footprints.

\subsubsection{Energy Consumption and Modeling Methodology}
Accurate power estimation requires integrating detailed power models into cycle-accurate simulators.

\vspace{5pt}
\noindent
\emph{Dynamic and Leakage Power:}
Dynamic Power arises from transistor switching activity during logic operations and memory access.
Leakage Power represents static loss during idle states.
In NVM-based PIM, leakage power is typically much lower than DRAM, allowing aggressive power-gating techniques \cite{Dong2012NVSim}.

\vspace{5pt}
\noindent
\emph{Energy Efficiency (TOPS/W):}
Tera Operations Per Second per Watt (TOPS) is used for measuring the efficiency of AI hardware accelerators. 
It allows fair comparison across systems with varying data types and operational scales by normalizing performance against power consumption \cite{Zhu2023MNSIM2.0}.

\subsubsection{Physical and Economic Constraints}
Beyond energy, architectural feasibility also depends on physical layout and cost.

\vspace{5pt}
\noindent
\emph{Area:}
Silicon area is measured in mm\(^2\) and includes core logic, interconnects, and memory arrays. It often scales with the fabrication technology node (\textit{F}). Tools like \texttt{CACTI} \cite{Balasubramonian2017CACTI} provide joint estimates of area, timing, and leakage to support design space exploration with trade-offs between performance density and cost \cite{Dong2012NVSim}.

\subsection{Thermal and Reliability Metrics}
To ensure long-term operational correctness and safety, modern simulators must also model temperature dynamics and component-level reliability. These factors are especially crucial in 3D-stacked and high-density memory systems like PIM.

\subsubsection{Reliability and Lifetime}
Reliability encompasses the ability of memory elements and compute units to maintain correct operation over time.
This metric is measured based on various parameters, as outlined below.

\vspace{5pt}
\noindent
\emph{Endurance (Write Cycles):}
Endurance is measured by the maximum number of program/erase or write cycles a memory cell can undergo before it becomes unreliable. This is particularly important for NVM technologies like Phase-Change Memory (PCM) and Resistive RAM (ReRAM), where limited endurance can restrict applicability in high-write environments \cite{Dong2012NVSim, wong2010phase}.

\vspace{5pt}
\noindent
\emph{Data Retention:}
This metric refers to the duration for which a memory cell can reliably hold data without refresh. While DRAM requires frequent refresh due to its volatile nature, NVM technologies offer significantly longer data retention, making them suitable for persistent memory applications \cite{Dong2012NVSim}.

\vspace{5pt}
\noindent
\emph{Process Variation and Non-Ideal Factors:}
At the device level, simulators model the effects of
$a)$ manufacturing variability (e.g., Gaussian resistance noise or threshold voltage); $b)$ permanent defects (stuck-at-faults). These non-idealities are particularly relevant in analog or low-precision digital computation, as they can introduce significant functional errors in PIM circuits \cite{shafiee2016isaac}.

\subsubsection{Thermal Stability and Functional Accuracy}
Thermal management ensures the physical stability of the chip, while accuracy verifies the correct functionality of the resulting computation.

\vspace{5pt}
\noindent
\emph{Thermal Modeling:}
Thermal simulations track heat dissipation and peak operating temperatures across memory chips. In DRAM, elevated temperatures require more frequent refresh operations, reducing overall performance and increasing energy use. Tools like DRAMSim3 \cite{Li2020DRAMsim3} and HotSpot \cite{han2021thermal} are often used together to co-model access scheduling and thermal behavior.

\vspace{5pt}
\noindent
\emph{Functional Accuracy:}
For AI accelerators, particularly those executing Convolutional Neural Networks (CNNs), functional accuracy becomes a critical metric. Since PIM hardware may use low-precision data types or analog computation (which introduces quantization errors and noise), simulators must include full training or inference workflows to assess the impact of hardware constraints on model accuracy.


\section{Benchmarking Methodologies for PIM Architectures}
\label{sec:sim_benchmarks}

In the realm of PIM architectures, benchmarks serve as a critical foundation for research and evaluation. They provide a standardized measurement framework that allows researchers to evaluate and compare different designs, algorithms, and simulators under standardized uniform conditions.
Without such standardized benchmarks, studies would resort to disparate workloads, inconsistent metrics, and varying inputs, making cross-comparison and reproducibility nearly impossible. 
Benchmarks offer several critical benefits such as:

\begin{itemize}
    \item \textit{Reproducibility}, shared workload ensure experiments can be repeated and results independently validated.
    \item \textit{Comparability}, by evaluating architectures on the same benchmark set, results gain semantic meaning, allowing for fair, side-by-side comparisons across papers or research efforts.
    \item \textit{Stress Testing}, well-designed benchmarks expose architectural strengths and weaknesses, whether they are memory-bound, compute-bound, or suffer from irregular access patterns, thus guiding design improvements.
\end{itemize}

In short, benchmarks act as both reference workloads and performance standards that tie together otherwise fragmented research into a coherent, comparable discourse.

\subsection{Types of Benchmarks}
From a functional perspective, PIM benchmarks fall into two broad categories: \textit{micro-level} or \textit{architectural-level} benchmarks.

\subsubsection{Micro-Level Benchmarks}
These are fine-grained that evaluate small, isolated functions such as memory copy, vector addition, or arithmetic reduction. They are ideal for measuring core metrics like latency, bandwidth, and energy consumption of specific architectural components. They are lightweight and well-suited for validating correctness or testing specific architectural features; however, they do not reflect the complex interactions present in the end-to-end applications.

\subsubsection{Architectural-Level Benchmarks}
These benchmarks include complete workloads, synthetic traces, or high-level applications (e.g., neural networks, graph analytics). They evaluate the behavior of the entire system, including memory hierarchy, compute patterns, interconnect behavior, and scalability. 
As such, they are essential for assessing end-to-end performance and energy efficiency in realistic scenarios.

Benchmarks may also be classified by their origin and specialization into two categories; \textit{general} and \textit{PIM-specific} benchmarks.
\subsubsection{General Benchmarks}
 General benchmarks are usually derived from traditional CPU or GPU suites, such as SPEC~\cite{bucek2018spec}, Phoenix~\cite{git2016phoenix}, Rodinia~\cite{che2009rodinia}, or GAP~\cite{beamer2015gap}.
 These are partially modified to offload specific kernels onto PIM hardware.
While they provide familiar workloads, they may not fully capture PIM-specific execution characteristics.
\subsubsection{PIM-specific Benchmarks}
Designed from the ground up to target PIM architectures, these benchmarks emphasize memory-centric computations, including bitwise logic, near-bank reductions, or matrix–vector operations executed directly within memory arrays. These workloads are essential for uncovering behaviors unique to PIM systems.

\subsection{Simulator-Integrated Benchmarking Environments}
Several simulators integrate built-in benchmark suites to support reproducible and low-level exploration of PIM behavior.

\subsubsection{\texttt{PIMSim} and \texttt{PiMulator}}
These simulators provide lightweight, self-contained workloads that users can execute directly for validating functionality or exploring core performance metrics~\cite{xu2018pimsim, git2022PiMulator}.

\texttt{PIMSim} includes a built-in suite of vector and logic kernels (e.g., addition, copy, bitwise operations) and a few graph-oriented routines like BFS and PageRank, based on the PEI benchmark set. \texttt{PiMulator} extends this concept by integrating the Hopscotch microbenchmark suite along with native PIM primitives such as RowClone, LISA, and Ambit.
This enables users to experiment with near-bank and in-DRAM compute patterns.

A few simulators deliver more comprehensive benchmark collections that model real applications rather than synthetic kernels. \texttt{uPIMulator} is tightly coupled with the PrIM benchmark suite, providing ready-to-run workloads such as GEMV, SpMV, histogram, and database filtering directly within its framework~\cite{Hyun2024uPIMulator}.
\texttt{PIMeval}, on the other hand, supports the full \texttt{PIMbench} benchmark set, which includes 16 diverse tasks ranging from logical operations to machine-learning workloads, making it one of the most complete toolchains for evaluating digital DRAM PIM architectures~\cite{10763591}.
\texttt{MNSIM 2.0} incorporates end-to-end simulation of neural network workloads such as LeNet, VGG, and ResNet under varying quantization and process variation conditions~\cite{Zhu2023MNSIM2.0}. 
\texttt{MemTorch} provides a modular interface for converting standard PyTorch models into memristive PIM workloads, allowing evaluation of both analog and digital compute-in-memory accelerators without additional integration effort~\cite{git2022MemTorch}.

Some simulators such as \texttt{Ramulator-PIM}, \texttt{MultiPIM}, and \texttt{MPU-Sim} rely entirely on external benchmark suites (e.g., Rodinia, GAP, CUDA kernels) used for validation studies~\cite{git2025ramulatorPIM, Yu2021MultiPIM, Xie2022MPUSim}.
While these demonstrate architectural feasibility, they lack standardized internal benchmarking environments, reducing reproducibility.
As a result, only a subset of simulators provide integrated, reusable benchmarking environments that support consistent, in-depth exploration of PIM systems.

\subsubsection{Dedicated Benchmark Suites for PIM}
Recognizing the need for standardization, several benchmark suites have been developed specifically to support PIM research. The most prominent among them are \texttt{InSituBench}, \texttt{PrIM}, and \texttt{PIMbench}.

\subsubsection{InSituBench Benchmark}
\texttt{InSituBench} was proposed as part of the Fulcrum framework, which aims to simplify the design and evaluation of in-situ accelerators, systems that perform computation directly inside memory arrays~\cite{lenjani2020fulcrum}. The core motivation was  to address the inconsistency in workloads used by researchers, which made fair comparisons across PIM architectures difficult.

Unlike general-purpose CPU or GPU benchmarks, \texttt{InSituBench} focuses on the specific operational characteristics of in-memory and near-data processing, such as limited compute granularity, data locality, and reduced off-chip communication.

\texttt{InSituBench} contains a diverse mix of workloads chosen to represent the most common access and computation patterns seen in real PIM use cases. These workloads are grouped into five categories~\cite{lenjani2020fulcrum}:

\begin{enumerate}
    \item \textit{Bitwise Operations}: Basic logical primitives (AND, OR, XOR, NOT) used in PIM logic operations and bit-serial arithmetic.
    \item \textit{Arithmetic Kernels}: Integer and floating-point addition and multiplication operations, typical in reduction, histogramming, and filtering tasks.
    \item \textit{Data Movement and Reduction}: Includes memory copy, initialization, and aggregation kernels (sum, max), focusing on minimizing data transfer.
    \item \textit{Vector and Matrix Workloads}: GEMV, SpMV, and Conv2D kernels simulate ML and signal processing tasks.
    \item \textit{Graph and Search Tasks}: BFS, PageRank, and histogram traversal, designed to stress irregular memory access and fine-grained control.
\end{enumerate}

Each workload operates on an abstract model built around the Fulcrum's dispatch mechanism, which defines a simple yet flexible interface for dispatching in-situ operations within a memory stack. This enables the same benchmarks to be executed under different architectural configurations allowing fair comparisons across platforms. The authors validate the suite on both FPGA-based prototypes and cycle-level simulators, demonstrating how it captures the trade-offs between flexibility, parallelism, and control overhead in real in-situ architectures~\cite{lenjani2020fulcrum}.

In summary, \texttt{InSituBench} establishes a comprehensive, PIM-specific benchmarking framework covering low-level bitwise operations, arithmetic computation, data-movement primitives, and higher-level analytical kernels. Its modular structure allows designers to isolate the effects of hardware organization, communication hierarchy, and control granularity. By bridging low-level kernel evaluation with application-inspired workloads, \texttt{InSituBench} provides a practical and reproducible foundation for comparing emerging in-situ and PIM architectures.

\subsubsection{PrIM}
\texttt{Processing-In-Memory benchmarks(PrIM)} is the first comprehensive benchmark suite designed specifically for evaluating commercial PIM hardware, particularly the UPMEM's  architecture~\cite{gomez2021benchmarking}. \texttt{PrIM} provides an open-source set of workloads that cover a wide range of memory-bound applications. Unlike synthetic microbenchmarks or simulation-only workloads, \texttt{PrIM} enables researchers to study real performance, scalability, and energy efficiency compared to CPUs and GPUs.

The benchmark suite includes 16 workloads, each selected from diverse domains such as dense and sparse linear algebra, databases, analytics, graph processing, neural networks, bioinformatics, and image processing~\cite{gomez2021benchmarking}. The workloads are categorized as follows:

\begin{enumerate}
    \item \textit{Linear Algebra:} GEMV, SpMV, and Vector Addition.
    \item \textit{Databases:} Filtering and uniqueness checks.
    \item \textit{Analytics:} Binary search and time-series analysis.
    \item \textit{Graph Algorithms:} BFS and traversal tasks.
    \item \textit{Neural Networks:} MLP inference under constrained memory.
    \item \textit{Bioinformatics:} Needleman–Wunsch alignment.
    \item \textit{Image Processing:} Histogram kernels.
    \item \textit{Parallel Primitives:} Reduction, scan, and transpose operations.
\end{enumerate}

\texttt{PrIM}’s key strength lies in its well-balanced combination of workload diversity and practical relevance.
The suite encompasses a broad spectrum of applications, ranging from simple data-parallel patterns such as vector addition and reduction to complex, synchronization-intensive tasks like BFS and Needleman–Wunsch (NW). This extensive coverage effectively reflects the complete computational range encountered in memory-bound workloads. Each benchmark within \texttt{PrIM} has been rigorously characterized using the roofline performance model, confirming that performance is genuinely constrained by memory access bandwidth rather than by compute throughput~\cite{gomez2021benchmarking}.

The authors evaluated \texttt{PrIM} on two different UPMEM PIM systems—one with 640 DPUs and another with 2,556 DPUs—and compared the results against those from state-of-the-art CPU and GPU platforms. Their findings demonstrated that PIM architectures can achieve substantial performance improvements, with speedups of up to 23× and energy efficiency gains of approximately 5×. These benefits are particularly pronounced for workloads that are heavily memory-bound, involve relatively simple arithmetic operations, and exhibit minimal inter-DPU communication overhead~\cite{gomez2021benchmarking}. In essence, \texttt{PrIM} serves a dual purpose: it functions both as a rigorous scientific benchmark suite and as a diagnostic tool that identifies which classes of workloads are most amenable to PIM acceleration. Additionally, it provides valuable guidance for co-design efforts in both hardware and software aimed at optimizing future memory-centric computing architectures.

\subsubsection{PIMbench}
\texttt{PIMbench} is part of the \texttt{PIMeval/PIMbench} framework, a unified platform designed for modeling, simulating, and benchmarking of digital DRAM-based PIM architectures~\cite{10763591}. This framework was developed to address a significant gap in the PIM research community: the lack of a portable and standardized benchmark suite applicable across wide variety of PIM designs including bit-serial, subarray-level, and bank-level architectures. Unlike suites benchmarks such as \texttt{PrIM} or \texttt{InSituBench}, which are tied to specific hardware implementations, \texttt{PIMbench} provides a high-level PIM programming API that allows the same benchmark code to execute unmodified on different architectural platforms, promoting reproducibility and enabling fair cross-platform comparison.

\texttt{PIMbench} contains 16 benchmarks covering a wide range of computation and data access patterns, categorized into multiple domains~\cite{10763591}:
\begin{enumerate}
    \item \textit{Linear Algebra}: includes Vector Addition, AXPY, GEMV, and GEMM designed to evaluate streaming memory access and arithmetic throughput.
    \item \textit{Sorting and Cryptography}: Radix Sort evaluates combined PIM–host execution performance, while AES encryption and decryption benchmarks measure logical intensity and bitwise computation.
    \item \textit{Graph and Database}: Triangle Count and Filter-by-Key benchmarks evaluate random access and associative processing capabilities.
    \item \textit{Image Processing}: includes Histogram, Brightness Adjustment, and Image Downsampling, which capture data aggregation and pixel-level computation.
    \item \textit{Machine Learning}: K-Nearest Neighbors (KNN), Linear Regression, K-means clustering, and three convolutional neural networks (VGG-13, VGG-16, VGG-19) model both supervised and unsupervised learning workloads under varying arithmetic and communication demands.
\end{enumerate}

Each workload specifies memory access pattern (sequential vs. random) and execution type (PIM-only or PIM + Host), thereby emphasizing how data layout and host interaction affect overall system performance.
Collectively, these benchmarks span from simple computational kernels (like addition or reduction) to more complex hybrid tasks that combine in-memory and host-side execution~\cite{10763591}.

\texttt{PIMeval} simulator is evaluated across three digital DRAM-based PIM architectures: (1) subarray-level bit-serial, (2) Fulcrum-style bit-parallel, and (3) bank-level PIM. Their results show that bit-serial PIM excels at massively parallel logical and reduction operations, Fulcrum’s bit-parallel design performs best on arithmetic-heavy kernels, and bank-level PIM is advantageous for workloads requiring broader data access~\cite{10763591}. Overall, \texttt{PIMbench} facilitates quantitative, cross-architecture, quantitative evaluation of performance, energy efficiency, and scalability.
In summary, \texttt{PIMbench} represents a major advancement toward establishing a standardized benchmarking methodology within the PIM community. By combining a diverse range of workloads with a portable API interface, it effectively bridges the gap between architectural simulation and system-level analysis, promoting reproducibility and accelerating research in memory-centric computing.





\section{Comparative Study of PIM Simulators}
\label{sec:comparative_study}
To better understand the capabilities and limitations of current Processing-in-Memory (PIM) simulation tools, we performed a comparative study using two representative open-source simulators: \texttt{PIMeval} \cite{git2025PIMevalPIMbench} and \texttt{PIMSimulator} \cite{git2025PIMSimulator}.
Both tools aim to provide quantitative insights into PIM architectures, but they differ substantially in modeling depth, configuration flexibility, and output granularity.
We follow two goals in this study: 
\begin{itemize}
    \item Evaluating how well these frameworks model the performance of common key workloads.
    \item Understanding the modeling assumptions and simulation methodologies behind each tool.
\end{itemize}

\subsection{Motivation and workload selection}
Matrix multiplication kernels form the computational backbone of nearly all modern AI workloads. In both the training and inference phases of deep neural networks, large-scale linear algebra operations, particularly General Matrix–Matrix Multiplication (GEMM) and General Matrix–Vector Multiplication (GEMV), dominate runtime, memory bandwidth consumption, and overall energy cost.

For instance, in Convolutional Neural Networks (CNNs), convolution operations are typically transformed into GEMM kernels through tensor reshaping, making matrix–matrix multiplication the central compute primitive. In Transformer-based architectures, GEMM operations appear extensively in both the multi-head attention projections (query, key, and value computations) and the feed-forward layers, while GEMV arises in autoregressive decoding or token-by-token inference, where a single embedding vector is multiplied by large model weight matrices.

As Processing-in-Memory (PIM) architectures are explicitly designed to minimize data movement and accelerate such dense linear algebra operations, GEMM and GEMV represent the most meaningful and representative workloads for evaluating PIM simulators.

We selected four representative workloads that vary both in data dimensionality and operation type: 
\begin{enumerate}
    \item $512 \times 512$ in $512 \times 512$ matrix--matrix multiplication
    \item $1024 \times 1024$ in $1024 \times 1024$ matrix--matrix multiplication
    \item $512 \times 512$ in $512 \times 1$ matrix--vector multiplication
    \item $1024 \times 1024$ in $1024 \times 1$ matrix--vector multiplication
\end{enumerate}

Together, these workloads capture both the compute- and memory-intensive ends of AI workloads, allowing us to evaluate how each simulator models parallelism, bandwidth utilization, and latency scaling.

\subsection{Evaluated simulators}
\subsubsection{PIMeval/PIMbench}
PIMeval is a detailed simulation and benchmarking framework for PIM-based architectures. It models the internal structure of memory devices at the rank, bank, and subarray levels, while also capturing the hierarchical organization of PIM cores distributed across the DRAM.

The simulator focuses primarily on timing and bandwidth modeling.
It estimates key performance characteristics such as row activation latency, read/write latency, column access time, and achievable bandwidth.
These fine-grained models enable PIMeval to predict the latency and throughput of large-scale memory operations under various architectural configurations. 

PIMeval supports a range of memory configurations, including DDR, HBM, LPDDR, and GDDR, as shown in Table~\ref{table2}. While the original implementation described in the base paper focused solely on DDR, additional memory technologies have been introduced in the public GitHub repository. However, not all of these are fully implemented—e.g., HBM currently lacks support for bit-serial execution.

\begin{table}[htbp]
\centering
\caption{PIMeval configuration example}
\label{table2}
\begin{tabular}{|l|c|}
\hline
\textbf{Parameter} & \textbf{Value} \\
\hline
Number of Ranks & 1 \\
Banks per Rank & 128 \\
Subarrays per Bank & 32 \\
Rows per Subarray & 8192 \\
Columns per Subarray & 8192 \\
Number of PIM Cores & 2048 \\
Simulation Target & Bank-level PIM device \\
\hline
\end{tabular}
\end{table}
For our experiment, we use (BANK\_LEVEL), FULCRUM, and bit-serial (BITSERIAL) architectures.  In the bank-level architecture, operations are directly mapped to DRAM banks, which makes it simple, energy efficient, and fast. On the other hand, FULCRUM, the Fulcrum architecture features parallelism at the sub-array level inside memory and more flexible dataflow, and BITSERIAL is a fully digital bit-serial PIM execution style, often optimized for logical operations and faces high area and time overhead for arithmetic operations.

PIMeval reports the following output metrics:
\begin{itemize}
\item \textit{Operation Latency (ms):} total delay for read, write, compute, and data movement operations
\item \textit{Bandwidth Utilization (GB/s):} achieved throughput compared to the theoretical peak
\item \textit{Energy per Operation (mJ):} estimated dynamic and static energy consumption 
\item \textit{Datamovement's latency (ms) and energy (mJ):} estimated delay and consumed energy during data transfers between memory and processing units
\end{itemize}
These outputs make PIMeval a timing-accurate, architecture-oriented simulator, ideal for exploring how DRAM-level organization affects PIM performance.

\subsubsection{PIMSimulator}
PIMSimulator provides a modular, instruction-driven simulation framework designed to capture the logical and functional behavior of PIM-enabled DRAMs (based on HBM interface).
Rather than focusing solely on physical timing, it models the execution flow of PIM instructions—such as addition, multiplication, reduction, and data movement—across distributed PIM cores.

This simulator supports a wide range of architectural configurations, including: 
\begin{itemize}
    \item Number of channels, ranks, and banks
    \item PIM core allocation across subarrays
    \item Memory read/write latency models
    \item Host–PIM communication overhead
    \item Custom PIM instruction sets and scheduling policy
\end{itemize}
Example rows of PIMSimulator hardware configuration used in our experiments are shown in Table~\ref{table3}.

\begin{table}[htbp]
\centering
\caption{PIMSimulator configuration example}
\label{table3}
\begin{tabular}{|l|c|}
\hline
\textbf{Parameter} & \textbf{Value} \\
\hline
Number of Bank Groups & 4 \\
Number of PIM Blocks & 4 \\
Device Width & 8 \\
Burst Length (BL) & 8 \\
Read Latency (RL) & 19 ns\\
Row-to-Row Delay (tRRDL) & 7 ns\\
Row-to-Column Delay for READ (tRCDRD) & 19 ns \\
Write Recovery Time (tWR) & 20 ns\\
Clock Cycle Time (tCK) & 0.75 ns\\
Power-Down Current – Precharge Power-Down Mode (IDD2P) & 34 mA\\
Core Voltage (Vddc) & 1.2 V\\
Wordline Boost Voltage (Vpp) & 2.5 V\\
\hline
\end{tabular}
\end{table}

PIMSimulator’s outputs are more abstract than PIMeval’s, focusing on operation counts, execution time, and memory transactions. This makes it suitable for exploring algorithmic mapping strategies (e.g., distributing GEMM tiles across PIM cores) or evaluating different instruction scheduling schemes. However, PIMeval is more realistic and considers energy overhead. 

\subsection{Results and Discussions}
Table \ref{tab:perf} summarizes the execution performance for all configurations. PIMeval demonstrated significantly higher performance in the FULCRUM architecture compared to Bank-level, highlighting the efficiency of subarrays in facilitating parallelism. Bit-serial architecture is slower per arithmetic operation, which is why we experience throughput degradation in this fine-grained bit-level model.

For the 1024×1024 GEMM, PIMeval’s BANK LEVEL mode achieved 408.2 GOPS, compared to 251.5 GOPS in FULCRUM and 64.7 K GOPS in the BITSERIAL mode. The 512×512 GEMM followed the same trend, scaling roughly with matrix size. In GEMV, performance dropped sharply, since vector operations expose less parallelism; the BANK LEVEL mode achieved 0.21 GOPS for the 1024×1024 GEMV, compared to 0.05 GOPS for PIMSimulator.

\begin{figure}[h]
\centering
\includegraphics[scale=0.3]{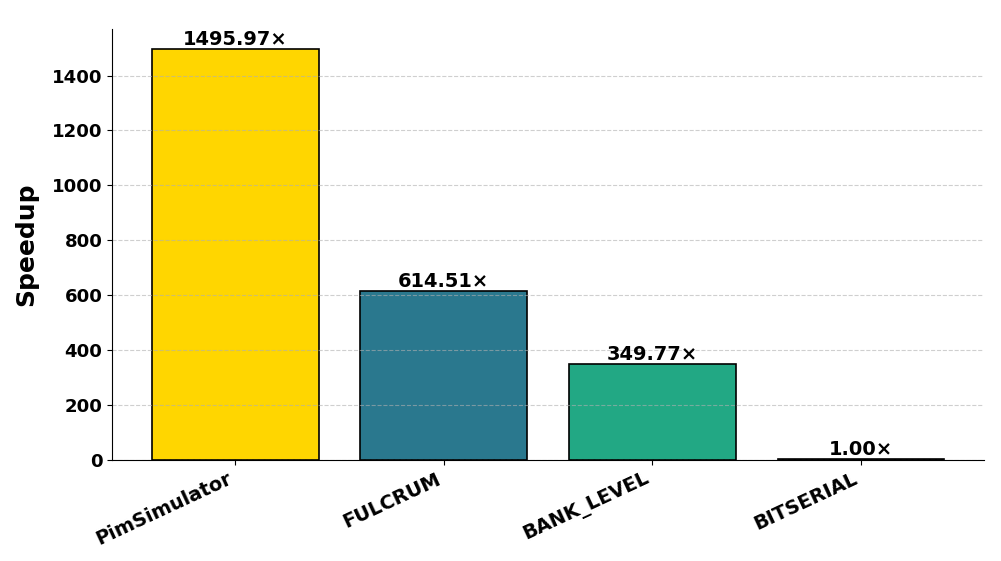}
\caption{Speedup comparisson in PIMeval and PIMSimulator for $512 \times 512$ GEMV workload: PIMSimulator does not consider energy overhead, and totally focuses on optimizing latency and performance.}\label{fig:Gemv_512}
\end{figure}

\begin{figure}[h]
\centering
\includegraphics[scale=0.3]{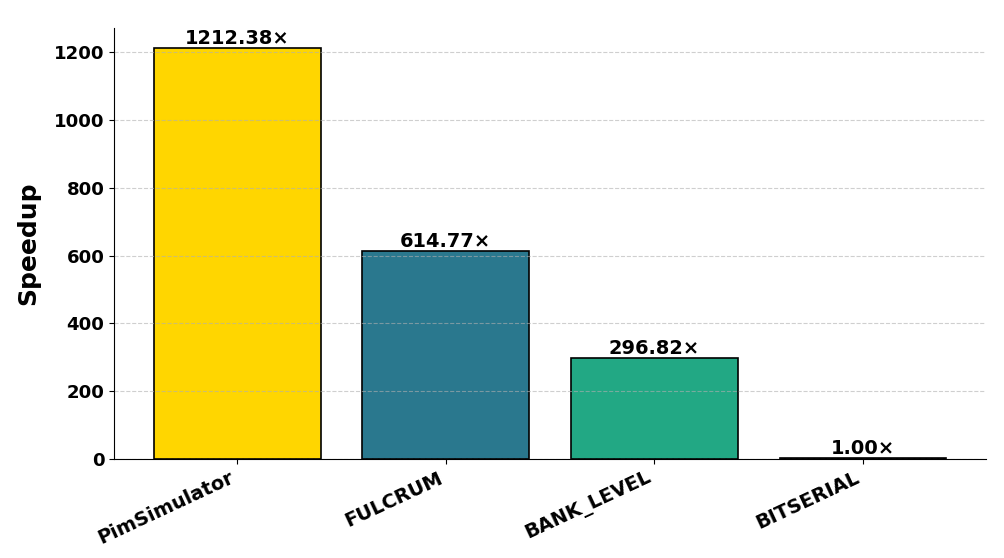}
\caption{Speedup comparisson in PIMeval and PIMSimulator for $1024 \times 1024$ GEMV workload.}\label{fig:Gemv_1024}
\end{figure}

PIMSimulator consistently reports lower execution time compared to PIMeval as shown in Figures~\ref{fig:Gemv_512} and~\ref{fig:Gemv_1024}, as it mainly focuses on optimizing scheduling and performance without considering energy efficiency. 

In terms of energy, the Bank-Level configuration is consistently the most energy-efficient.
For 1024×1024 GEMM, it consumes 569 mJ, compared to 3.8 J for FULCRUM and >1 kJ for Bit-Serial.
A similar pattern appears in GEMV workloads, where Bit-Serial energy grows disproportionately with matrix size.

\begin{table*}[t]
\centering
\caption{Performance and Energy Comparison of PIMSimulator and PIMeval under GEMM/GEMV Workloads}
\label{tab:perf}
\resizebox{\textwidth}{!}{
\begin{tabular}{|l|cccc|}
\hline
\textbf{Workload / Metric} &
\textbf{PIMSimulator} &
\textbf{PIMeval} &
\textbf{PIMeval} &
\textbf{PIMeval} \\
 & & \textbf{(Bank-Level)} & \textbf{(FULCRUM)} & \textbf{(BITSERIAL)} \\
\hline
\textbf{1024$\times$1024 GEMM (ms)} & -- & 408.23 & 251.49 & 64680.94 \\
\textbf{1024$\times$1024 GEMV (ms)} & 0.0521 & 0.2128 & 0.1027 & 63.16 \\
\textbf{512$\times$512 GEMM (ms)} & -- & 88.31 & 62.88 & 16170.64 \\
\textbf{512$\times$512 GEMV (ms)} & 0.0211 & 0.0903 & 0.0514 & 31.58 \\
\hline
\textbf{1024$\times$1024 GEMM (mJ)} & -- & 569.40 & 3831.23 & 1011350.62 \\
\textbf{1024$\times$1024 GEMV (mJ)} & -- & 0.3359 & 1.5006 & 987.65 \\
\textbf{512$\times$512 GEMM (mJ)} & -- & 129.09 & 957.99 & 237223.47 \\
\textbf{512$\times$512 GEMV (mJ)} & -- & 0.1524 & 0.7507 & 463.33 \\
\hline
\end{tabular}
}
\end{table*}

\clearpage
\section{Conclusion}
\label{sec:conclusion}
Processing-in-Memory (PIM) has matured from a conceptual paradigm into a central research direction for overcoming the memory wall in modern computing systems. By reducing data movement and integrating computation within memory structures, PIM offers a promising path toward higher energy efficiency and parallelism. However, this architectural shift also introduces unprecedented challenges in system modeling, workload characterization, and performance validation. Simulation has thus become the most indispensable tool in this evolution, providing the analytical foundation to explore, refine, and validate PIM concepts long before physical realization.

Over the past decade, simulation frameworks have evolved significantly, moving from simple memory latency models to sophisticated multi-level environments that capture detailed timing, architectural behavior, and application workloads. The diversity of simulators reflects the multidimensional nature of PIM itself: each framework balances accuracy, scalability, and flexibility in a distinct way. Some emphasize high-level abstraction to enable rapid exploration, while others offer cycle-accurate fidelity for microarchitectural verification. Together, these complementary approaches form an ecosystem that supports the entire design pipeline, from device characterization to algorithm evaluation.

A major outcome of this exploration is the growing emphasis on consistency and comparability. As PIM research expands across academic and industrial domains, standardized methodologies and benchmarks have become crucial for credible evaluation. The availability of open-source simulators and shared workloads has fostered reproducibility and accelerated progress, allowing diverse research groups to validate findings on common grounds. This trend highlights the collaborative nature of the PIM community and the pivotal role of simulation in sustaining it.

Looking ahead, the future of PIM simulation lies in hybrid and intelligent modeling. Adaptive frameworks that combine fast functional simulation with selective high-fidelity modeling can provide both scalability and precision. The integration of machine learning techniques promises to automate design-space exploration, enabling simulators to predict performance trends and optimize configurations dynamically. Moreover, tighter coupling between simulation and prototype hardware will bridge the gap between modeling and measurement, ensuring that simulation results remain grounded in physical reality.

Simulation is not merely a supporting tool but the driving engine of PIM innovation. It enables the translation of architectural imagination into verifiable insight and accelerates the path toward practical deployment. The continued refinement of simulation methodologies will determine how rapidly and reliably PIM technologies transition from research laboratories to mainstream computing systems.

\clearpage

  \bibliographystyle{elsarticle-num} 
  \bibliography{references}



%
%
%
\end{document}